\documentclass[12pt,a4paper]{article}
\usepackage{graphicx}
\usepackage{color}
\usepackage{bm}
\usepackage{cite}
\usepackage{epsfig}
\usepackage{amsmath,amssymb}
\usepackage{booktabs}
\makeatletter

\usepackage{caption}
\allowdisplaybreaks[1]
\definecolor{nicered}{rgb}{0.9,0.1,0.1}
\definecolor{nicegreen}{rgb}{0.3,0.8,0.1}
\definecolor{niceblue}{rgb}{0.2,0.1,0.8}
\usepackage[colorlinks=true, linkcolor=niceblue, citecolor=niceblue, urlcolor=niceblue]{hyperref} 

\begin{document}

\title{SUSY threshold corrections to quark and lepton mixing \\inspired by $SO(10)$ GUT models}
\author{
  \centerline{
    Yu~Muramatsu$^{1}$\footnote{E-mail address: yumurasub@gmail.com}
    ~and 
    Yoshihiro~Shigekami$^{2}$\footnote{E-mail address: sigekami@post.kek.jp}}
  \\[25pt]
  \centerline{
    \begin{minipage}{\linewidth}
      \begin{center}
        $^1${\it \normalsize Institute~of~Particle~Physics~and~Key~Laboratory~of~Quark~and~Lepton~
Physics~(MOE), Central~China~Normal~University,~Wuhan,~Hubei~430079,
~People's~Republic~of~China} \\
        $^2${\it \normalsize 
        School~of~physics, Huazhong~University~of~Science~and~Technology, \\
        Wuhan,~Hubei~430074,
        ~People's~Republic~of~China} 
      \end{center}
  \end{minipage}}
  \\[40pt]}
\date{}
\maketitle

\begin{abstract}
%In grand unified theories (GUTs), two types of unification are predicted. 

In the standard model (SM), the Cabibbo-Kobayashi-Maskawa (CKM) matrix is the only origin of flavor violations (FVs). 
On the other hand, in general, there are a lot of sources of the FVs in a physics beyond the SM, through the diagonalizing matrices which make Yukawa matrices diagonal. 
Although most of the diagonalizing matrices are unknown ones, grand unified theories (GUTs) can fix these matrices.
In particular, if we consider a GUT model based on the $SO(10)$ group, these diagonalizing matrices are strongly related to each other because of the matter unification.
However, this unification causes the problem on the realization of measured SM fermion masses and mixing angles.
Up to now, many people showed that supersymmetric (SUSY) threshold corrections can solve this problem, especially unfavorable fermion mass relations. 
In this paper, we show that how these SUSY threshold corrections affect unknown diagonalizing matrices. 
Since these contributions are also related to the FVs induced by SUSY particles, we can estimate the maximal effect to the diagonalizing matrices.
We found that the $(1, 3)$ and $(3, 1)$ elements of diagonalizing matrices for quarks can be as large as the corresponding mixing angle of the CKM matrix. Moreover, we found that the higher SUSY scale can predict the larger mixing angles, although strong cancellation between the Higgs mass parameters is needed to realize the electroweak symmetry breaking.
\end{abstract}

\newpage

\section{Introduction}
\label{sec:intro}
Observing Higgs boson \cite{Aad:2012tfa} establishes the standard model (SM) completely.
However, some topics such as non-zero neutrino masses and existence of the dark matter are not explained in the SM, and then, many physics beyond the SM (bSMs) are established to explain these topics.
So far, many experiments have tried to find the signals of the new physics and they reported the results about upper and/or lower limits or observed values for some observables.
We can now test the bSMs by using these limits and observables.

One of the most important tools to test the bSMs is flavor phenomena.
Flavor observables can probe effects of new particles with masses far above the energy scale of current collider experiments.
In the SM, the only origin of flavor violations (FVs) is the Cabibbo-Kobayashi-Maskawa (CKM) matrix \cite{CKMmatrix} which is defined by a product of the mixing matrices of left-handed quarks. 
Small mixing angles of the CKM matrix cause small FVs for the SM predictions, and therefore, flavor observables show us clear signal of the bSMs.

In general, there are additional origins of the FVs in the bSMs.
Especially, the diagonalizing matrices play a crucial role among these origins.
Well-known examples are supersymmetric (SUSY) models.
In these models, FVs can be induced by mediating the SUSY particles.
To predict flavor observables via contributions from SUSY particles, mass insertion parameters (MIPs) in a super-CKM basis \cite{Hall:1985dx} are often used. 
The super-CKM basis is a basis in which the rotations for obtaining diagonal Yukawa matrices are applied to sfermions as well as the SM fermions. 
Because of this rotation, MIPs include information of the corresponding diagonalizing matrices.
However, most of the diagonalizing matrices are unknown ones because what we know about these matrices from the experiments is only the CKM matrix \footnote{Although the Pontecorvo-Maki-Nakagawa-Sakata (PMNS) matrix is also the experimental observable, we focus only on the correction to the CKM matrix in this paper. This is because the corrections considered in this paper are expected to be small, and then, such small corrections are irrelevant for the PMNS matrix which has large mixing angles.}.

One of the possibilities to know the information of these unknown diagonalizing matrices is considering the model with grand unifications.
In grand unified theories (GUTs), the three gauge interactions in the SM are unified into a single one, and moreover, the particles in the SM are unified into fewer multiplets \cite{Georgi:1974sy}.
The latter unification reduces degrees of freedom of the diagonalizing matrices.
In the GUTs, new particles are predicted, and they also induce FVs through unknown diagonalizing matrices.
Although these contributions are negligible for most processes because new particles which induce FVs have masses around the GUT scale, 
%However, 
proton decays have a sensitivity for these contributions.
Recently, the Super-Kamiokande Collaboration reported some events which can be candidates for proton decay, and these include a candidate for FV decay of the proton such as $p\to \pi \mu$ decay mode \cite{SKFVPC}.
It has been confirmed that in some models, this FV decay of the proton can be comparable to main decay mode of the proton \cite{FVPD}.

In the minimal $SO(10)$ GUT models \cite{minSO10} \footnote{In this paper, we define the models in which only 16-dimensional representation is used for quarks and leptons as minimal $SO(10)$ GUT models. In addition, we use only one 10-dimensional representation for two Higgses in SUSY models.}, all quarks and leptons are unified into only one multiplet ${\bf 16}$ for one family.
Because of this unification, all diagonalizing matrices become the CKM-like matrix to realize the CKM matrix except for a light neutrino mixing matrix.
Note that the mixing angles of the light neutrino mixing matrix can be large because of new degrees of freedom from seesaw mechanism \cite{seesawI, seesawII}.
However because of the unification, it is not easy to realize measured quark and lepton masses and mixings. 
One useful correction to solve this problem is a SUSY threshold correction \cite{Hall:1985dx, SUSYthreshold}. 

These SUSY threshold corrections mainly come from the soft SUSY-breaking trilinear couplings $A_{\psi}$ (A-terms) and the Higgsino mass parameter $\mu$, and are especially used to realize measured quark and lepton masses \cite{YU_SUSYthreshold}.
It is known that off-diagonal parts of the A-terms contribute to the CKM matrix \cite{SUSYthresholdCKM}.
This means that the SUSY threshold corrections also contribute to the diagonalizing matrices.
Therefore, it is important to estimate the contributions from the SUSY threshold correction to the diagonalizing matrices.

In this paper, we discuss how large diagonalizing matrices originated from the SUSY threshold corrections are allowed in the minimal SUSY $SO(10)$ GUT model. 
In this setup, we may obtain some specific signal for the model from the effect of large diagonalizing matrices. 
Note that large off-diagonal parts of fermion mass matrices which generate large diagonalizing matrices come from large off-diagonal parts of the A-terms through the SUSY threshold corrections.
%The effect of large diagonalizing matrices may  a specific signal for the model. 
However, these off-diagonal parts are strongly constrained by flavor phenomena like SUSY flavor-changing neutral currents (FCNCs) \footnote{The effects of these chirally enhanced corrections on FCNC processes have been studied, for example in Ref.~\cite{Crivellin:2009ar}.}. 
Furthermore, these also cause the negative mass squared for sfermions through the renormalization group equation (RGE) effects. 
Then, we calculate the maximal size of diagonalizing matrices numerically without conflicting with these concerns. 
%Therefore in this paper, we discuss how large diagonalizing matrices are allowed under these constraints.

This paper is organized as follow. In Sec.~\ref{sec:notation}, we show the notation used in this paper.
We give the method of our calculation in Sec.~\ref{sec:calculation}, and 
our main results are shown in Sec.~\ref{sec:results}. 
The conclusion is in Sec.~\ref{sec:conclusion}. 
In addition, we show the detail of the SUSY threshold corrections in Appendix~\ref{appendix:SUSYthreshold}. 
We also summarize the relevant constraints to our calculation
in Appendix~\ref{appendix:ConstMIP}.
The discussion about the maximal SUSY threshold correction which satisfy the constraints are shown in Appendix~\ref{appendix:RM_SUSYthreshold}.

\section{Notation}
\label{sec:notation}
In this section, we summarize our notation in this paper.
The superpotential for the Yukawa interactions and $\mu$ term in the interaction basis is
\begin{equation}
W= \epsilon_{ab} \left[ (Y_u)_{ij} Q_{L\, i}^a U_{R\, j}^c H_u^b - (Y_d)_{ij} Q_{L\, i}^a D_{R\, j}^c H_d^b - (Y_e)_{ij} L_{L\, i}^a E_{R\, j}^c H_d^b \right] + \epsilon_{ab} \mu H_u^a H_d^b,
\end{equation}
where $Q_{L}$, $L_{L}$, $U_{R}^c$, $D_{R}^c$, $E_{R}^c$, $H_u$, and $H_d$ are superfields of the left-handed quark doublet, left-handed lepton doublet, right-handed up-type quark, right-handed down-type quark, right-handed charged-lepton, up-type Higgs doublet, and down-type Higgs doublet, respectively.
In this paper, we use the capital letters for superfields and the small letters for the corresponding SM components.
Here, $Y_u$, $Y_d$, and $Y_e$ are the up-type quark, down-type quark, and charged lepton Yukawa matrices, $a$, $b = 1,\, 2$ are weak isospin indices, $i$, $j=1,\, 2,\, 3 $ are generation indices, and $\epsilon_{12} \equiv +1$ is an antisymmetric tensor.
To obtain fermion masses, neutral components of CP-even Higgses develop vacuum expectation values (VEVs) $\langle h^0_{u(d)} \rangle \equiv v_{u(d)}$ whose ratio is defined as $\tan \beta \equiv v_u / v_d$.
The mass basis of fermions are obtained by diagonalizing Yukawa matrices as
\begin{equation}
\psi_{L\, i} (Y_\psi)_{ij} \psi_{R\, j}^c = (L_\psi^{\dagger} \psi_L)_i (L_\psi^{T} Y_\psi R_\psi)_{ij} (R_\psi^{\dagger} \psi_R^c)_j = (\hat{Y}_\psi)_{ii} \psi'_{L\, i} \psi'{}_{R\, i}^c,
\end{equation}
where $\psi$ and $\psi'$ indicate the SM fermion in the interaction basis and mass basis, $L_{\psi}$ and $R_{\psi}$ are diagonalizing matrices for the left-handed and right-handed fermion, and $(\hat{Y}_\psi)_{ii}$ is the Yukawa coupling of $i$-th generation fermion $\psi$. 
The same rotations are applied to the corresponding sfermions because of the SUSY, and this basis is called the super-CKM basis. 
In this paper, $\hat{X}$ refers to the parameter of $X$ in the super-CKM basis.
We calculate mixing angles for the diagonalizing matrices, defined as 
\begin{align}
 L_{\psi}^T &\equiv
  \begin{pmatrix}
   c_{12}^{\psi L} & -s_{12}^{\psi L} & 0 \\
   s_{12}^{\psi L*} & c_{12}^{\psi L} & 0 \\
   0 & 0 & 1
  \end{pmatrix}
  \begin{pmatrix}
   c_{13}^{\psi L} & 0 & -s_{13}^{\psi L} \\
   0 & 1 & 0 \\
   s_{13}^{\psi L*} & 0 & c_{13}^{\psi L}
  \end{pmatrix}
  \begin{pmatrix}
   1 & 0 & 0 \\
   0 & c_{23}^{\psi L} & -s_{23}^{\psi L} \\
   0 & s_{23}^{\psi L*} & c_{23}^{\psi L}
  \end{pmatrix}, 
\label{eq:mixing angleL}\\
 R_{\psi} &\equiv
 \begin{pmatrix}
  1 & 0 & 0 \\
  0 & c_{23}^{\psi R} & s_{23}^{\psi R} \\
  0 & -s_{23}^{\psi R*} & c_{23}^{\psi R}
 \end{pmatrix}
 \begin{pmatrix}
  c_{13}^{\psi R} & 0 & s_{13}^{\psi R} \\
  0 & 1 & 0 \\
  -s_{13}^{\psi R*} & 0 & c_{13}^{\psi R}
 \end{pmatrix}
 \begin{pmatrix}
  c_{12}^{\psi R} & s_{12}^{\psi R} & 0 \\
  -s_{12}^{\psi R*} & c_{12}^{\psi R} & 0 \\
  0 & 0 & 1
 \end{pmatrix},
\label{eq:mixing angleR}
\end{align}
where $s_{ij}^{\psi L/R} \equiv \sin\theta_{ij}^{\psi L/R}e^{-i\chi_{ij}^{\psi L/R}}$, and 
$c_{ij}^{\psi L/R} \equiv \cos\theta_{ij}^{\psi L/R}$ \footnote{For simplicity, we do not consider CP phases $\chi_{ij}^{\psi L/R}$ in this paper.}.

To calculate the SUSY threshold corrections, we use following soft-breaking parameters in a Lagrangian $\mathcal{L}_{\text{soft}}$,
\begin{eqnarray}
\mathcal{L}_{\text{soft}} &=& -\frac{1}{2} \left( M_1 \tilde{b} \tilde{b} + M_2 \tilde{w} \tilde{w} + M_3 \tilde{g} \tilde{g} \right) - \epsilon_{ab} B_\mu h_u^a h_d^a + h.c. \nonumber \\
&&- \epsilon_{ab} \left[ (A_u)_{ij} \tilde{q}_{L\, i}^a \tilde{u}^*_{R\, j} h_u^b - (A_d)_{ij} \tilde{q}_{L\, i}^a \tilde{d}^*_{R\, j} h_d^b - (A_e)_{ij} \tilde{l}_{L\, i}^a \tilde{e}^*_{R\, j} h_d^b \right] + h.c. \nonumber \\
&& - m_{h_u}^2 h_u^\dagger h_u - m_{h_d}^2 h_d^\dagger h_d - (m_{\tilde{q}}^2)_{ij} \tilde{q}^\dagger_{L\, i} \tilde{q}_{L\, j}- (m_{\tilde{l}}^2)_{ij} \tilde{l}^\dagger_{L\, i} \tilde{l}_{L\, j} \nonumber \\
&& - (m_{\tilde{u}}^2)_{ij} \tilde{u}^*_{R\, i} \tilde{u}_{R\, j} - (m_{\tilde{d}}^2)_{ij} \tilde{d}^*_{R\, i} \tilde{d}_{R\, j} - (m_{\tilde{e}}^2)_{ij} \tilde{e}^*_{R\, i} \tilde{e}_{R\, j}
\end{eqnarray}
where $M_i$ ($i = 1, 2, 3$) are gaugino masses, and $\tilde{\psi}$ is the SUSY partner for the corresponding SM particle $\psi$.
Moreover, we define following 6$\times$6 sfermion mass squared matrices as 
\begin{equation}
\mathcal{L}_{\text{mass}} = - \hat{M}_{\tilde{\psi}\, ij}^2 \tilde{\psi}_i^\dagger \tilde{\psi}_j, \quad \tilde{\psi}=\begin{pmatrix}
\tilde{\psi}_{L\, 1} & \tilde{\psi}_{L\, 2} & \tilde{\psi}_{L\, 3} & \tilde{\psi}_{R\, 1} & \tilde{\psi}_{R\, 2} & \tilde{\psi}_{R\, 3}
\end{pmatrix}^T,
\end{equation} 
where
\begin{equation}
\hat{M}_{\tilde{u}}^2 = \begin{pmatrix}
\hat{m}^2_{\tilde{q}} + v_u^2 \hat{Y}_u \hat{Y}_u & v_u \hat{A}_u^* - \mu v_d \hat{Y}_u \\
v_u \hat{A}_u^T - \mu^* v_d \hat{Y}_u & \hat{m}^2_{\tilde{u}} + v_u^2 \hat{Y}_u \hat{Y}_u
\end{pmatrix}\equiv \begin{pmatrix}
\Delta_{LL}^u & \Delta_{LR}^u \\
\Delta_{RL}^u & \Delta_{RR}^u
\end{pmatrix},
\label{eq:up sfermion mass2}
\end{equation}
\begin{equation}
\hat{M}_{\tilde{d}}^2 = \begin{pmatrix}
V_{\text{CKM}}^\dagger \hat{m}^2_{\tilde{q}} V_{\text{CKM}} + v_d^2 \hat{Y}_d \hat{Y}_d & v_d \hat{A}_d^* - \mu v_u \hat{Y}_d \\
v_d \hat{A}_d^T - \mu^* v_u \hat{Y}_d & \hat{m}^2_{\tilde{d}} + v_d^2 \hat{Y}_d \hat{Y}_d
\end{pmatrix}\equiv \begin{pmatrix}
\Delta_{LL}^d & \Delta_{LR}^d \\
\Delta_{RL}^d & \Delta_{RR}^d
\end{pmatrix},
\label{eq:down sfermion mass2}
\end{equation}
\begin{equation}
\hat{M}_{\tilde{e}}^2 = \begin{pmatrix}
\hat{m}^2_{\tilde{l}} + v_d^2 \hat{Y}_e \hat{Y}_e & v_d \hat{A}_e^* - \mu v_u \hat{Y}_e \\
v_d \hat{A}_e^T - \mu^* v_u \hat{Y}_e & \hat{m}^2_{\tilde{e}} + v_d^2 \hat{Y}_e \hat{Y}_e
\end{pmatrix}\equiv \begin{pmatrix}
\Delta_{LL}^e & \Delta_{LR}^e \\
\Delta_{RL}^e & \Delta_{RR}^e
\end{pmatrix}.
\label{eq:electron sfermion mass2}
\end{equation}
The sfermion mass squared matrix $\hat{M}_{\tilde{\psi}\, ij}^2$ is diagonalized by a unitary matrix $U_{\tilde{\psi}}$ as 
\begin{equation}
\hat{M}_{\tilde{\psi}, \text{diag}}^2=U_{\tilde{\psi}} \hat{M}_{\tilde{\psi}}^2 U_{\tilde{\psi}}^\dagger.
\end{equation}

The SUSY threshold corrections are induced by chirality-flipping self-energies.
In Ref.~\cite{Crivellin:2011jt}, 
the details of the calculation for these corrections are shown in a decoupling limit $M_{\text{SUSY}}\gg v$ where $M_{\text{SUSY}}$ and $v$ are the SUSY scale and the VEV of the SM Higgs. 
We also apply to this decoupling limit in this paper. 
In the Appendix~\ref{appendix:SUSYthreshold}, we summarize the SUSY threshold corrections in our notation.

The SUSY threshold corrections affect on the diagonalizing matrices if chirality-flipping self-energies have off-diagonal parts which come from the off-diagonal parts of A-terms and sfermion mass squared matrices. 
The size of these off-diagonal parts are important to obtain the large SUSY threshold corrections.
However, these off-diagonal parts are also related to the SUSY FCNCs, and therefore, these sizes are constrained from the experimental measurements.
Moreover, large off-diagonal parts of A-terms make the sfermion squared mass negative and cause the a charge and color breaking (CCB) \cite{CCB}. 
%In general, these constraints are discussed by using the MIPs, defined as
%\begin{eqnarray}
%( \delta_{AB}^{\psi} )_{ij} \equiv \frac{( \Delta_{AB}^{\psi} )_{ij}}{\sqrt{\hat{M}_{\tilde{\psi} \, ii}^2 \hat{M}_{\tilde{\psi} \, jj}^2}},
%\label{eq:MIpara}
%\end{eqnarray}
%where $A, B = L$ or $R$, $\psi = u, d, e$. 
%$( \Delta_{AB}^{\psi} )_{ij}$ is matrix element of $6 \times 6$ sfermion mass matrix defined in Eqs.~\eqref{eq:up sfermion mass2}, \eqref{eq:down sfermion mass2}, and \eqref{eq:electron sfermion mass2}.
In the Appendix~\ref{appendix:ConstMIP}, we summarize the SUSY FCNC and CCB constraints, used in this paper.

\section{Method of the calculation}
\label{sec:calculation}
From the following, we explain the calculation of the diagonalizing matrices with the SUSY threshold corrections.
To calculate the diagonalizing matrices, we have to set the input parameters at the GUT scale. 
In our model, these parameters are the unified gauge coupling $g_{\text{GUT}}$, unified Yukawa matrix $Y_{\text{GUT}}$, unified gaugino mass $M_{1/2}$, the scale of A-terms $A_0$, universal sfermion mass $m_{16}^2$, Higgs mass parameter $m_{10}^2$, and Higgs mass difference $\Delta m_h^2$ which is needed to realize the electroweak symmetry breaking (EWSB).
Note that for simplicity and for reducing the parameters, we assume $\Delta m_h^2 \sim M_{\rm SUSY}^2$.
We discuss the Yukawa unification at the beginning \footnote{In this paper, we only discuss the Yukawa unification for third generation fermions. For the first and second generations fermions, the observed fermion masses can be achieved by non-renormalizable interactions \cite{Ellis:1979fg} due to the smallness of couplings compared with those of third generation fermions.} since we can fix part of the input parameters from it.

\subsection{SUSY threshold correction for realizing Yukawa unification}
The SUSY threshold correction plays an important role for realizing the Yukawa unification in the SUSY $SO(10)$ GUT models.
When we consider this Yukawa unification, we can fix the unified gauge coupling $g_{\text{GUT}}$ and (3,3) component of unified Yukawa matrix $(Y_{\text{GUT}})_{33}$. 
The latter one is obtained by setting the values of the SUSY threshold corrections and $\tan \beta$ appropriately. 
Note that some of the other parameters can be fixed by the other assumptions, as we explain later.

First of all, we find the required SUSY threshold correction and $\tan \beta$ for realizing the Yukawa unification. In this calculation, we start from the central values of the experimental measurements summarized in Table~\ref{tab:input}.
\begin{table}[tb]
\begin{center}
\begin{tabular}{|c|c||c|c|} \hline
$m_s/((m_u + m_d)/2)$ & $27.3^{+0.7}_{-1.3}$ & $m_e$ & $0.5109989461 (31)$ MeV \\
$m_d$ & $4.67^{+0.48}_{-0.17}$ MeV & $m_\mu$ & $105.6583745 (24)$ MeV \\
$m_s$ & $93^{+11}_{-5}$ MeV & $m_\tau$ & 1776.86$\pm 0.12$ MeV \\
$m_c$ & $1.27 \pm 0.02$ GeV & $M_Z$ & 91.1876(21) GeV \\%$M_H$ & 125.158$\pm 0.16$ GeV \\
$m_{b}(m_b)$ & 4.180$^{+0.04}_{-0.03}$ GeV & $M_W$ & $80.379(12)$ GeV \\ 
$m_t$(direct detection)& 173.0$\pm 0.4$ GeV & $G_F$ & 1.1663787(6)$\times 10^{-5}$ GeV$^{-2}$ \\
$\sin^2 \theta_W$ & 0.23122(4) & $\alpha_s(M_Z)$ & $0.1181(11)$ \\
$\alpha$ & 1/137.036 & & \\ \hline
\end{tabular}
\caption{The input parameters in our analysis. All values come from Ref.~\cite{PDG}.}
\label{tab:input}
\end{center}
\end{table}
There are three scales which are relevant to our calculation: $M_Z$, $M_{\rm SUSY}$, and the GUT scale $M_{\rm GUT}$.
First, we calculate the parameters at the $M_Z$ scale from the input parameters in Table~\ref{tab:input}.
For calculating the running quark masses, we use the Mathematica package RunDec \cite{RunDec}, while for the running lepton masses, we translate the lepton pole masses to running masses at the $M_Z$ scale, following Ref.~\cite{SMrun}. 
Next, we derive the couplings at the SUSY scale from those at the $M_Z$ scale, using the SM RGE at the two-loop level \cite{Machacek_Vaughn,Luo:2002ey} \footnote{
In some points, Ref.~\cite{Machacek_Vaughn} is inconsistent with Ref.~\cite{Luo:2002ey}. In this paper we use the RGE in Ref.~\cite{Luo:2002ey} which is consistent with Ref.~\cite{Bednyakov:2012en, Chetyrkin:2013wya}. Our calculation results at the SUSY scale are consistent with Ref.~\cite{Antusch:2013jca} which provides the SM Yukawa couplings around TeV scale in the $\overline{\rm MS}$ scheme and in the $\overline{\rm DR}$ schemes.}.
At the SUSY scale, we translate the $\overline{\text{MS}}$ scheme into the $\overline{\text{DR}}$ scheme according to Ref.~\cite{MStoDR}. 
After applying the matching condition by using the $\tan \beta$, we add the SUSY threshold corrections as
\begin{eqnarray}
y_t(M_{\text{SUSY}}) &=& \frac{1}{\sin \beta} y^{(0)}_t(M_{\text{SUSY}}) + \Delta y_t (M_{\text{SUSY}}), \nonumber \\
y_b(M_{\text{SUSY}}) &=& \frac{1}{\cos \beta} y^{(0)}_b(M_{\text{SUSY}}) + \Delta y_b (M_{\text{SUSY}}) , 
\label{eq:matching} \\
y_\tau(M_{\text{SUSY}}) &=& \frac{1}{\cos \beta} y^{(0)}_\tau(M_{\text{SUSY}}) + \Delta y_\tau(M_{\text{SUSY}}), \nonumber
\end{eqnarray}
where $\Delta y_\psi$ shows the SUSY threshold correction for the fermion $\psi$.
However in this section, we only consider $\Delta y_b$ because this contribution is dominant over the other ones.
Here $^{(0)}$ means parameters below the SUSY scale, namely, SM parameters.
Finally, we derive the couplings at the GUT scale from those at the SUSY scale, using the minimal SUSY SM (MSSM) RGE at the two-loop level \cite{MSSMrun}.
Note that in our calculation, $M_{\rm GUT}$ and $g_{\rm GUT}$ are defined as
\begin{equation}
g_1 (M_{\text{GUT}}) = g_2 (M_{\text{GUT}}) \equiv g_{\text{GUT}},
\label{eq:GUT_def}
\end{equation}
where %$g_1$ and $g_2$ is the $U(1)_Y$ and $SU(2)$ gauge coupling, and $g_1$ is the GUT normalized coupling.
$g_1$ is the GUT normalized gauge coupling for the $U(1)_Y$, and $g_2$ is the gauge coupling for the $SU(2)_L$.
By means of this procedure, we can obtain the required value of $\Delta y_b$ and $\tan \beta$ for realizing the Yukawa unification: $y_t = y_b = y_\tau$.
%\begin{table}[htb]
%\begin{center}
%\begin{tabular}{|c|c||c|c|} \hline
%$m_\tau$ & 1776.86$\pm 0.12$ MeV & $m_{b}(m_b)$&4.180$^{+0.04}_{-0.03}$ GeV \\ 
%$m_t$(direct detection)& 173.0$\pm 0.4$ GeV  & $M_H$ & 125.158$\pm 0.16$ GeV \\ 
%$M_Z$ & 91.1876(21) GeV & $M_W$ & $80.379(12)$ GeV   \\ 
%$\sin^2 \theta_W$ & 0.23122(4) &  $G_F$  & 1.1663787(6)$\times 10^{-5}$ GeV$^{-2}$ \\ 
%$\alpha$ & 1/137.036 & $\alpha_s(M_Z)$ & $0.1181(11)$ \\  \hline
%\end{tabular}
%\caption{The input parameters in our analysis. All values comes from Ref.~\cite{PDG}.}
%\label{tab:input}
%\end{center}
%\end{table}

\begin{figure}[tb]
\begin{center}
  \includegraphics[width=0.7\textwidth,bb= 0 0 360 229]{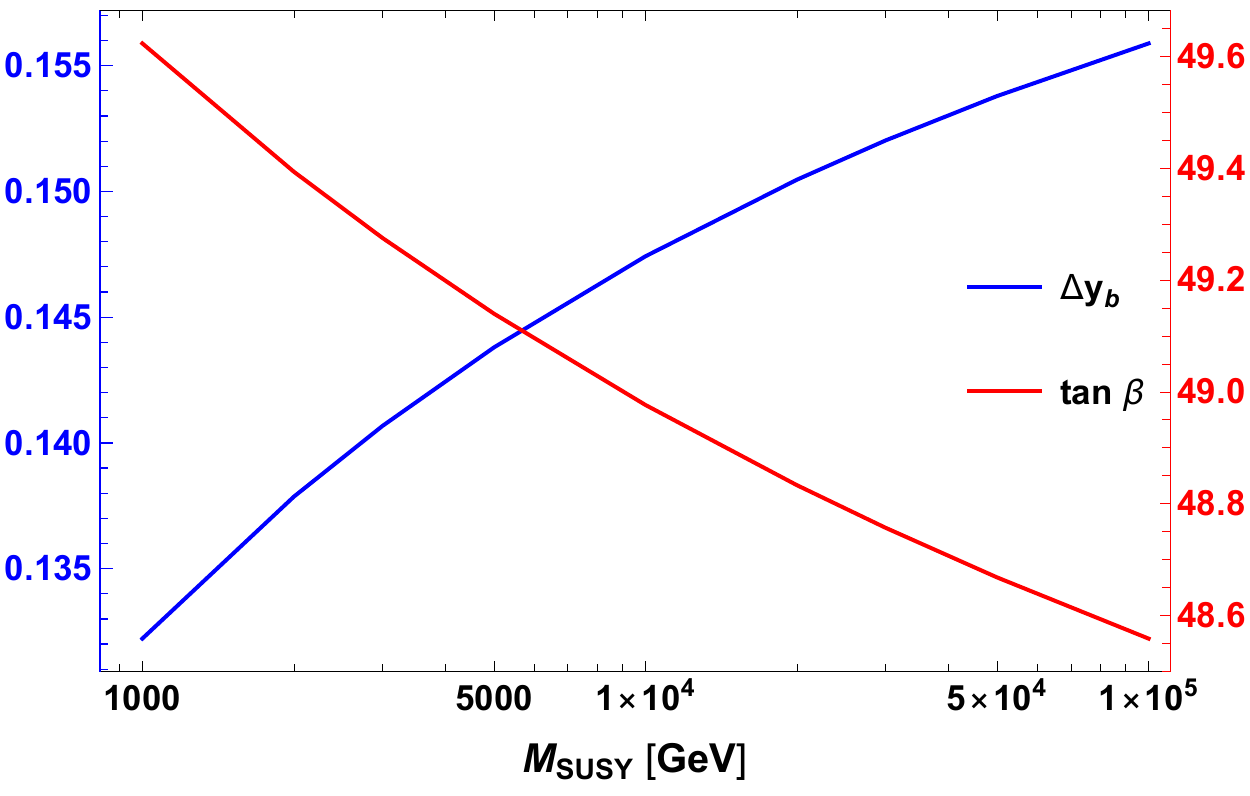}
\caption{Dependence of the required SUSY threshold correction $\Delta y_b$ and $\tan \beta$ for realizing the $SO(10)$ Yukawa unification on $M_{\text{SUSY}}$. The blue line and axis show the result for the required SUSY threshold correction, and the red ones show the result for the $\tan \beta$.}
\label{fig:req_SUSYthreshold}
\end{center}
\end{figure}

%Figure~\ref{fig:req_SUSYthreshold} shows the dependence of the required SUSY threshold correction $\Delta y_b$ and $\tan \beta$ to realize the Yukawa unification on the SUSY scale $M_{\text{SUSY}}$.
In Fig.~\ref{fig:req_SUSYthreshold}, we show the dependence of $\Delta y_b$ and $\tan \beta$ on $M_{\text{SUSY}}$.
%%%%%%%%%%%%%%%%%%%%%%%%%%%%%%%%%
%%%%%%%%%%%%%%%%%%%%%%%%%%%%%%%%%
The blue line and axis show the required SUSY threshold correction, and the red ones show the $\tan \beta$.
From this figure, importance of large $\tan \beta$ is clear which has been already discussed in Ref.~\cite{large_tanbeta}.
Interestingly, it is shown that a maximal SUSY threshold correction is larger than the required one because of the large $\mu$ contribution when the SUSY scale is below 100 TeV.
In Appendix~\ref{appendix:RM_SUSYthreshold}, we discuss this issue by considering the CCB and EWSB constraints. 
Note that the SUSY threshold correction for the diagonal part is dominated by the contributions from $\mu$ parameter in the case of large $\tan \beta$.
Therefore in this paper, the $\mu$ parameter is fixed to realize the required SUSY threshold correction, and we use a maximal A-term to maximize the effect to the diagonalizing matrices.

\subsection{SUSY threshold correction for the diagonalizing matrices}
\label{sec:SUSYthrefordiag}
In this section, we explain how to calculate the SUSY threshold correction for the diagonalizing matrices. 
In this calculation, we use the eigenvalues of Yukawa matrices at the GUT scale. % as parameters for the Yukawa matrices.
%As a result, without SUSY threshold corrections the Yukawa matrices are diagonal matrices at any energy scales, and the diagonalizing matrices are unit matrices.
%However this setup has advantages.
Therefore, when the SUSY threshold corrections are absent, the Yukawa matrices are diagonal matrices at any energy scales, and hence, the corresponding diagonalizing matrices are unit matrices.
There are some advantages in this setup.
First one is that contributions from the SUSY threshold corrections to the diagonalizing matrices become clear.
Origins of the SUSY threshold corrections to the diagonalizing matrices are off-diagonal parts of SUSY parameters $A_{\psi}$ and $m_{\tilde{\psi}}^2$ in the super-CKM basis.
%However, in the super-CKM basis, the off-diagonal parts of the SUSY parameters come from combination of these parameters in the interaction basis and the diagonalizing matrices.
%Therefore, contributions are unclear.
If the Yukawa matrices have the off-diagonal elements, these parts %in the super-CKM basis 
become the combination of those in the interaction basis and the diagonalizing matrices, which leads to the contributions are unclear.
The second advantage is that we can exclude some unfavorable contributions to the SUSY threshold corrections.
In many cases, family symmetries which show us how to realize the CKM matrix are introduced with the $SO(10)$ symmetry.
%Family symmetries show us how to realize the CKM matrices, and therefore in many cases family symmetries are introduced with the SO(10) symmetry.
However for this calculation, these family symmetries are unfavorable because these affect and fix the diagonalizing matrices.
By using diagonal Yukawa matrices, we can remove these unfavorable contributions.

In our calculation, we use the following assumptions at the GUT scale:
\begin{align}
&M_1 = M_2 = M_3 = M_{1/2}, \\[0.6ex]
&A_u = A_0 \hat{Y_u} + \delta A, ~~ A_d = A_0 \hat{Y_d} + \delta A, ~~ A_e = A_0 \hat{Y_e} + \delta A, \nonumber \\[0.6ex]
&m_{\tilde{q}}^2 = m_{\tilde{l}}^2 = m_{\tilde{u}}^2 = m_{\tilde{d}}^2 = m_{\tilde{e}}^2 = m_{16}^2 \, 1_{3 \times 3}, \\[0.6ex]
&m_{h_u}^2 = m_{10}^2 - \Delta m_h^2, ~~ m_{h_d}^2 = m_{10}^2 + \Delta m_h^2, \nonumber
\end{align}
where $\delta A$ indicates the off-diagonal part of the A-terms. Note that if $\delta A = 0$, each diagonalizing matrix becomes the unit matrix. 
Since we set $\Delta m_h^2 \sim M_{\rm SUSY}^2$, there are 5 parameters in this setup, $M_{1/2}$, $A_0$, $\delta A$, $m_{16}$, and $m_{10}$.
In order to reduce the parameters, we take the following assumptions: (i) the gluino mass at the SUSY scale is around $M_{\rm SUSY}$; (ii) the SUSY scale is defined by two stop masses as $M_{\rm SUSY} = \sqrt{m_{\tilde{t}_L}(M_{\text{SUSY}}) m_{\tilde{t}_R}(M_{\text{SUSY}})}$. 
Therefore, once we fix $A_0$, $\delta A$, and $m_{10}$ in each $M_{\rm SUSY}$ case, we can determine $M_{1/2}$ and $m_{16}$ by (i) and (ii), respectively. 
The remaining parameters can be determined so that the SUSY FCNC and CCB constraints are satisfied and the required SUSY threshold correction are obtained. 
Note that we also check whether the EWSB can be realized. 
Among such parameter sets, we take the set which maximizes the A-terms and calculate the diagonalizing matrices. 
%We show the input parameter at the GUT scale in Table~\ref{tab:input@GUTwithoff}.

After we calculate the diagonalizing matrices, we extract mixing angles by following Eqs.~\eqref{eq:mixing angleL} and \eqref{eq:mixing angleR}.
Moreover we compare these mixing angles and those of the measured CKM matrices \cite{PDG}, which are
%The mixing angles of the measured CKM matrices \cite{PDG} are 
\begin{equation}
s_{12}^{\text{CKM}}=2.2496 \times 10^{-1},\, s_{23}^{\text{CKM}}=4.16 \times 10^{-2},\, s_{13}^{\text{CKM}}=4.09 \times 10^{-3},
\end{equation}
where the CKM matrix $V_{\text{CKM}}$ is defined as
\begin{equation}
 V_{\text{CKM}} \equiv
 \begin{pmatrix}
  1 & 0 & 0 \\
  0 & c_{23}^{\text{CKM}} & s_{23}^{\text{CKM}} \\
  0 & -s_{23}^{\text{CKM}*} & c_{23}^{\text{CKM}}
 \end{pmatrix}
 \begin{pmatrix}
  c_{13}^{\text{CKM}} & 0 & s_{13}^{\text{CKM}} \\
  0 & 1 & 0 \\
  -s_{13}^{\text{CKM}*} & 0 & c_{13}^{\text{CKM}}
 \end{pmatrix}
 \begin{pmatrix}
  c_{12}^{\text{CKM}} & s_{12}^{\text{CKM}} & 0 \\
  -s_{12}^{\text{CKM}*} & c_{12}^{\text{CKM}} & 0 \\
  0 & 0 & 1
 \end{pmatrix}.
\end{equation}
%If the diagonalizing matrices are smaller than the CKM matrix, the SUSY threshold corrections to the CKM matrix can be negligible, and the CKM matrix can be obtained from the diagonalizing matrices come from Yukawa matrix. 
%are small, and the diagonalizing matrices come from Yukawa matrix structure to realize the CKM matrix.
If the diagonalizing matrices are as large as or larger than the CKM matrix, the SUSY threshold corrections affect on the CKM matrix.
In that case, the CKM matrix is obtained by the combination of the diagonalizing matrices from the SUSY threshold corrections and Yukawa matrices, or some cancellations are needed between the diagonalizing matrices to realize the CKM matrix.
%some diagonalizing matrices can be CKM-like matrix even if these are unit matrix without SUSY threshold corrections.
%If the diagonalizing matrices are larger than the CKM matrix, the CKM matix is realized by cancellation between the diagonalizing matrices, and the diagonalizing matrices can be large mixing matrix.

%\subsection{The procedure to find the maximal A-terms}

%To find the maximal A-terms, we adopt the following approach. First, we set the parameters at the GUT scale as a test value. Note that in our calculation, we set $\Delta m_h^2 = M_{\rm SUSY}^2$ and assume $A_{\psi} = A_0 \hat{Y}_{\psi} + \delta A$ where $A_0$ is the universal scale and $\delta A$ indicates the off-diagonal parts of the A-terms. Thanks to these assumption and the Yukawa unification, we have only 5 parameters ($M_{1/2},\, A_0,\, \delta A,\, m_{16}^2,\, m_{10}^2$) at the GUT scale. Second, we calculate the SUSY parameters at the SUSY scale by using parameters we set. Then, we compare the size of the off-diagonal parts of each A-term with the bounds on MIPs and CCB constraints. We also check whether the EWSB constraints which are discussed in Appendix \ref{appendix:RM_SUSYthreshold} are satisfied or not.

Since $s_{13}^{\rm CKM}$ is smaller than the other elements, we calculate $s_{13}^{\psi L, R}$ by setting $\delta A$ as
\begin{align}
\delta A = 
\begin{pmatrix}
0 & 0 & A_{\rm off} \\
0 & 0 & 0 \\
A_{\rm off} & 0 & 0
\end{pmatrix}.
\label{eq:deltaA}
\end{align}
In this case, we should take care about the SUSY FCNCs, especially $B_d$-$\overline{B_d}$ mixing and $\tau \to e \gamma$. 
The details of calculations and constraints for these processes are summarized in the Appendix~\ref{appendix:ConstMIP}. 
In the other case of $\delta A$, the contributions to the corresponding diagonalizing matrices are expected to be small due to the largeness of $s_{ij}^{\rm CKM}$ and/or the SUSY FCNC constraints.

\section{Results}
\label{sec:results}
The input parameters which satisfies the constraints and realizes the required SUSY threshold correction simultaneously are shown in Table~\ref{tab:input@GUTwithoff}.
\begin{table}[tb]
\begin{center}
\begin{tabular}{|c||c|c|c|} \hline
 & $M_{\rm SUSY}$ = 2 TeV & $M_{\rm SUSY}$ = 10 TeV & $M_{\rm SUSY}$ = 100 TeV \\ \hline \hline
%$\tan \beta (M_{\rm SUSY})$ & 49 & 49 & 49 \\ 
%$\Delta y_b (M_{\rm SUSY})$ & 0.14 & 0.15 & 0.16 \\ 
$M_{\rm GUT}$ & $1.2 \times 10^{16}$ GeV & $7.4 \times 10^{15}$ GeV & $3.6 \times 10^{15}$ GeV \\ 
$g_1 = g_2$ & 0.70 & 0.68 & 0.66 \\ 
$g_3$ & 0.70 & 0.69 & 0.67 \\ 
$y_u$ & $3.0 \times 10^{-6}$ & $3.0 \times 10^{-6}$ & $3.1 \times 10^{-6}$ \\ 
$y_c$ & $1.5 \times 10^{-3}$ & $1.5 \times 10^{-3}$ & $1.5 \times 10^{-3}$ \\ 
$y_d$ & $3.4 \times 10^{-4}$ & $3.4 \times 10^{-4}$ & $3.4 \times 10^{-4}$ \\ 
$y_s$ & $6.5 \times 10^{-3}$ & $6.5 \times 10^{-3}$ & $6.5 \times 10^{-3}$ \\ 
$y_e$ & $1.4 \times 10^{-4}$ & $1.3 \times 10^{-4}$ & $1.3 \times 10^{-4}$ \\ 
$y_\mu$ & $2.9 \times 10^{-2}$ & $2.8 \times 10^{-2}$ & $2.8 \times 10^{-2}$ \\ 
$y_t = y_b = y_\tau$ & 0.58 & 0.56 & 0.54 \\ 
$M_{1/2}$ & $9.3 \times 10^{2}$ GeV & $5.2 \times 10^{3}$ GeV & $5.9 \times 10^{4}$ GeV \\ 
$A_0^{\rm max}$ & $-5.5 \times 10^{3}$ GeV & $-2.4 \times 10^{4}$ GeV & $-1.8 \times 10^{5}$ GeV \\ 
$A_{\rm off}$ & $-98$ GeV & $-2.4 \times 10^{3}$ GeV & $-1.4 \times 10^{5}$ GeV \\ 
$m_{16}$ & $3.9 \times 10^{3}$ GeV & $1.7 \times 10^{4}$ GeV & $2.1 \times 10^{5}$ GeV \\ 
$m_{10}$ & $4.9 \times 10^{3}$ GeV & $2.0 \times 10^{4}$ GeV & $2.9 \times 10^{5}$ GeV \\ 
$\Delta m_h^2 / M_{\rm SUSY}^2$ & $1$ & $0.48$ & $1$ \\ \hline
\end{tabular}
\caption{The input parameters at the GUT scale. In this table, we show the results of the case where the structure of $\delta A$ is Eq.~\eqref{eq:deltaA}.}
%in which $(1,3)$ and $(3,1)$ elements of A-terms are present.}
\label{tab:input@GUTwithoff}
\end{center}
\end{table}
Here, $A_0^{\rm max}$ is the maximal value of $A_0$ which satisfies the CCB constraints. Note that the value of $\Delta m_h^2$ for $M_{\rm SUSY} = 10$ TeV should be smaller than $M_{\rm SUSY}^2$. We will explain this issue later. By using these input parameters, we can calculate the mixing angles for $L_{\psi}$ and $R_{\psi}$. %, and the results are summarized in Table~\ref{tab:s13results}. 
We show the results in Table~\ref{tab:s13results}. 
\begin{table}[tb]
\begin{center}
\begin{tabular}{|c|c|c|c|} \hline
$M_{\rm SUSY}$ & 2 TeV & 10 TeV & 100 TeV \\ \hline
$L_u$ & $-0.27$ & $-1.5$ & $-8.2$ \\
$R_u$ & $-0.25$ & $-1.4$ & $-7.8$ \\ \hline
$L_d$ & $0.27$ & $1.8$ & $9.6$ \\
$R_d$ & $0.54$ & $3.6$ & $17$ \\ \hline
$L_e$ & $-0.052$ & $-0.34$ & $-1.2$ \\
$R_e$ & $-0.12$ & $-0.81$ & $-2.8$ \\ \hline
\end{tabular}
\caption{The results of $s_{13}^{\psi L,R}$ in $10^{-3}$ unit for each case of $M_{\rm SUSY}$.}
\label{tab:s13results}
\end{center}
\end{table}
%Note that in each case, $c_{13}^{\psi L,R} \approx 1$ for all diagonalizing matrices. 
From these results, we found that $s_{13}^{q L, R}$ $(q = u, d)$ can be as large as $s_{13}^{\rm CKM}$ when $M_{\rm SUSY}$ becomes large. This is because the SUSY FCNC constraints become weak in large $M_{\rm SUSY}$ case, and therefore, we can take large $A_{\rm off}$. In this case, however, a strong cancellation between $| m_{h_u}^2 |$ and $m_{h_d}^2$ is needed to realize the EWSB. 
%Moreover, since the maximal SUSY threshold correction decreases as the SUSY scale increases (see Appendix~\ref{appendix:RM_SUSYthreshold}), 
%As explained in Sec.~\ref{sec:SUSYthrefordiag}, this mixing angle is originated from the off-diagonal parts of A-terms through the SUSY threshold corrections. Therefore, the size of $s_{ij}^{\psi L, R}$ is expected to be same order.

\begin{figure}[tb]
  \begin{center}
  \includegraphics[width=0.47\textwidth,bb= 0 0 450 307]{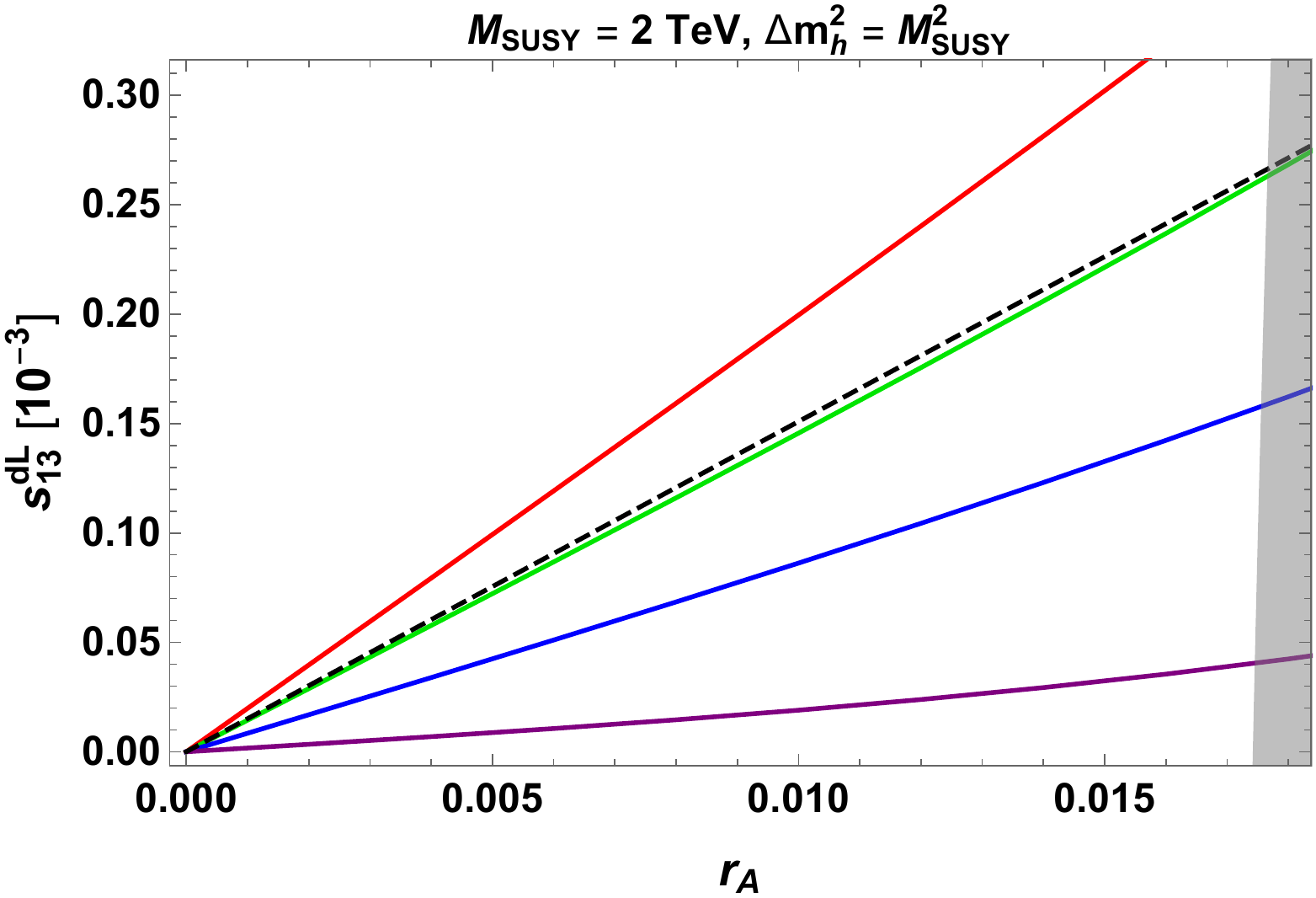} \hspace{0.3em}
  \includegraphics[width=0.5\textwidth,bb= 0 0 450 298]{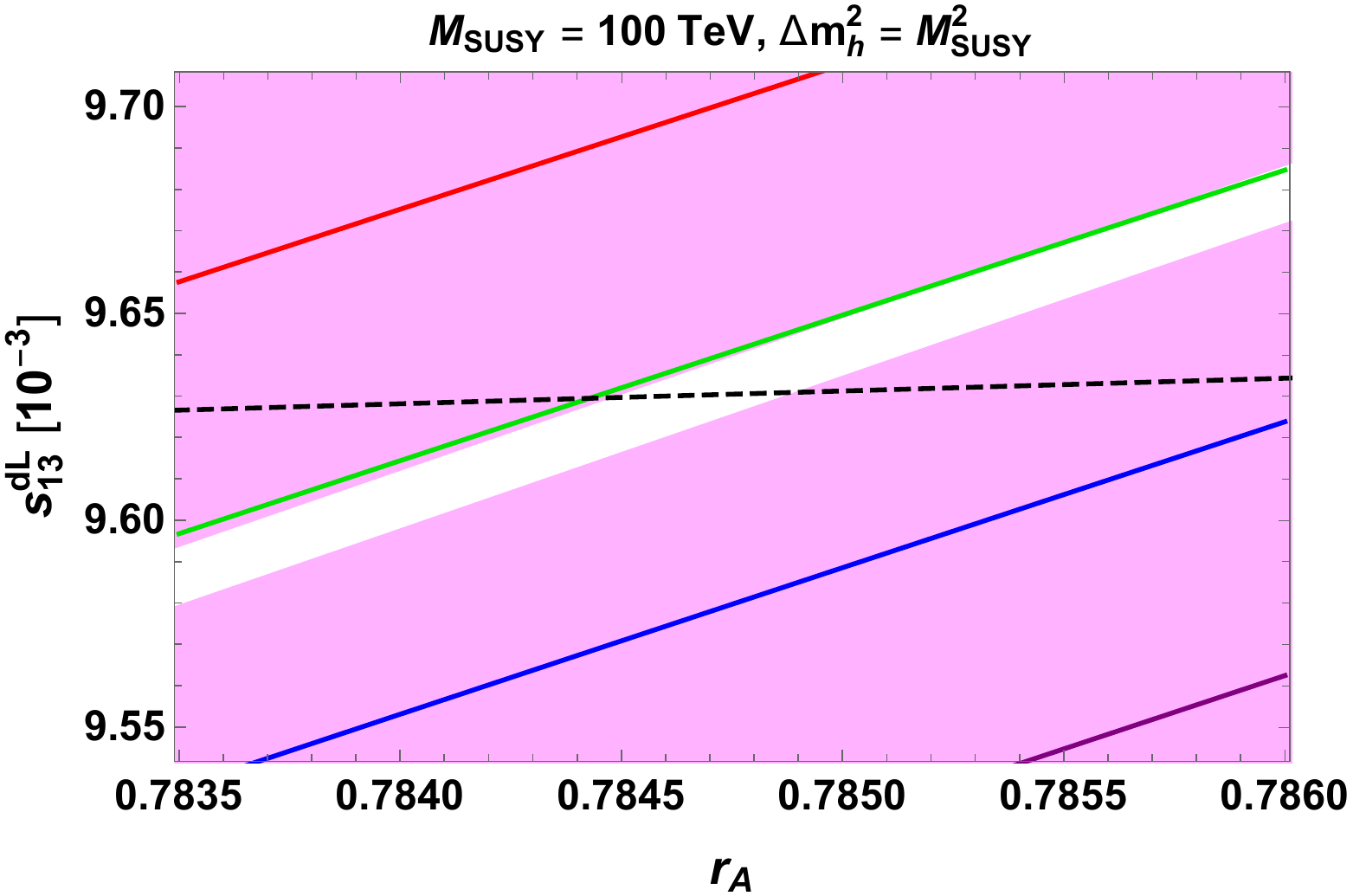}
  \caption{The parameter spaces allowed by the EWSB and SUSY FCNC constraints in $(r_A, s_{13}^{d L})$ plane. The magenta and gray shaded regions show the EWSB and SUSY FCNC constraints, respectively. The red, green, blue, and purple lines show the examples of prediction with different value of $m_{10}^2$, which correspond to $r_h = 1.6$, $1.605$, $1.61$, and $1.615$ in the left panel and $1.843$, $1.844$, $1.845$, and $1.846$ in the right panel. The black dashed line shows the relation between $s_{13}^{d L}$ and $r_h$ in each $r_A$, which realizes the required SUSY threshold correction.}
  \label{fig:parasp2and100}
  \end{center}
\end{figure}

\begin{figure}[tb]
  \begin{center}
    \includegraphics[width=0.6\textwidth,bb= 0 0 450 313]{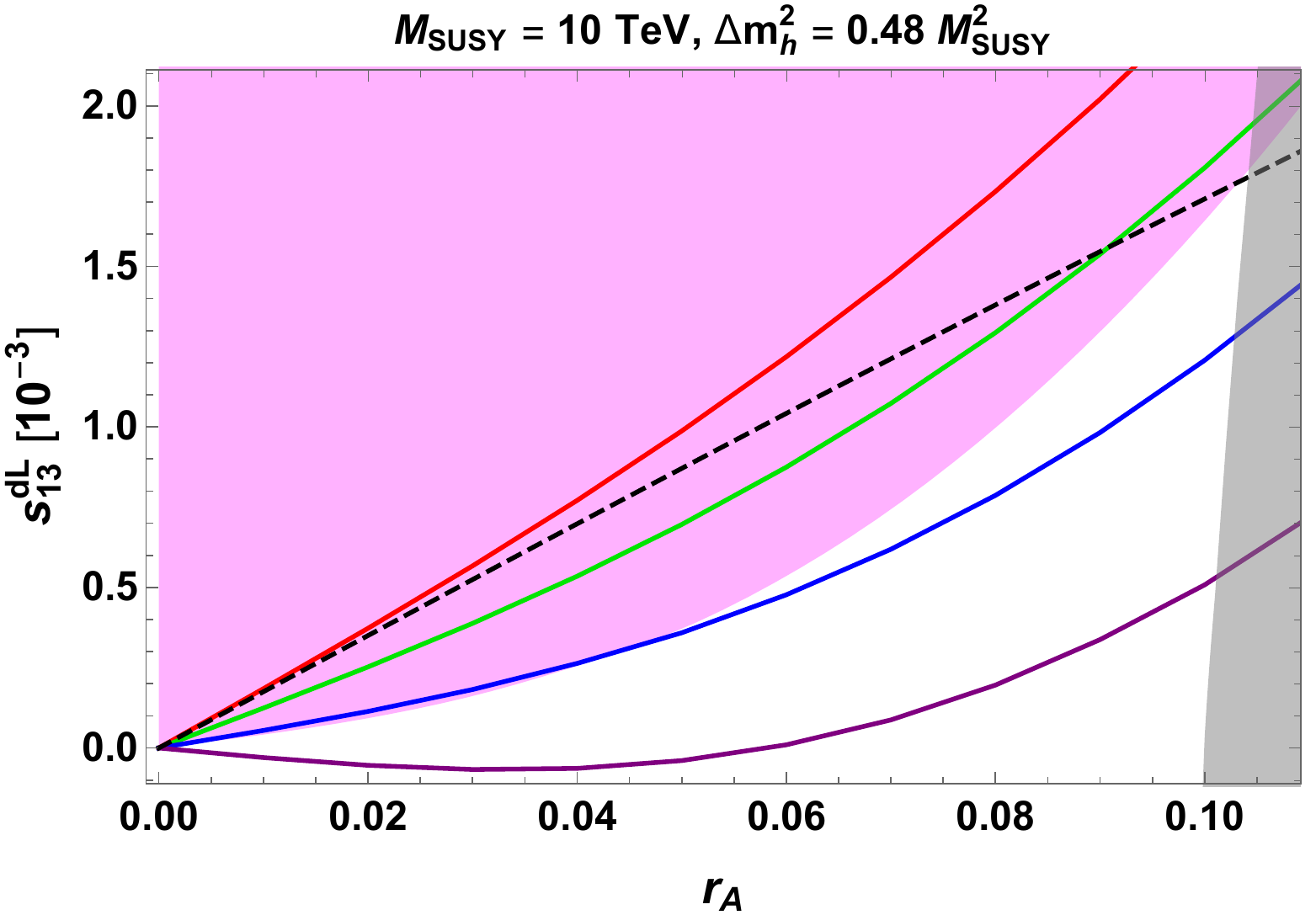}
  \caption{The parameter space allowed by the EWSB and SUSY FCNC constraints in $(r_A, s_{13}^{d L})$ plane. The color plots show the same meanings as in Fig.~\ref{fig:parasp2and100}. In this figure, the red, green, blue, and purple lines correspond to $r_h = 1.42$, $1.44$, $1.46$, and $1.48$, respectively.}
  \label{fig:parasp10}
  \end{center}
\end{figure}

%We found that the sizes of $s_{13}^L$ in $L_u$ and $L_d$ are comparable to $s_{13}^{\rm CKM}$ and these become large when the SUSY scale is large. 
%Moreover, $s_{13}^{L,R}$ for the charged-lepton are about one order of magnitude smaller than those for the quarks.
In order to specify that our input parameters realize the maximal size of $s_{13}^{\psi L, R}$ \footnote{The ``maximal size" of the mixing angle in this paper means the mixing angle which is calculated by the maximal value of the off-diagonal element of A-terms $A_{\rm off}$ through the SUSY threshold correction. Obviously, the maximal value of $A_{\rm off}$ can be calculated by the SUSY FCNC and the CCB constraints.}, we show the parameter spaces for $s_{13}^{d L}$ in each $M_{\rm SUSY}$ case in Figs.~\ref{fig:parasp2and100} and \ref{fig:parasp10}. 
We checked that the situation is same for the other mixing angles. The vertical axis is the mixing angles of $L_d$, and the horizontal one is the ratio of $A_{\rm off}$ and $A_0^{\rm max}$ at the GUT scale, which is denoted as $r_A$. The magenta and gray shaded regions show the EWSB and SUSY FCNC constraints, respectively. The red, green, blue, and purple lines show the examples of prediction with different value of $r_h \equiv m_{10}^2 / m_{16}^2$ at the GUT scale. The black dashed line indicates the relation between $s_{13}^{d L}$ and $r_h$ in each $r_A$, which realizes the required SUSY threshold correction for $y_b$. This means that the predicted $\Delta y_b$ is larger (smaller) than the required one above (below) the black dashed line. Note that we set $\Delta m_h^2 = M_{\rm SUSY}^2$ in Fig.~\ref{fig:parasp2and100}, while $\Delta m_h^2 = 0.48 M_{\rm SUSY}^2$ in Fig.~\ref{fig:parasp10}. This is because $\Delta m_h^2 < M_{\rm SUSY}^2$ is needed for the parameter space that satisfies the EWSB and SUSY FCNC constraints and realizes the required SUSY threshold correction simultaneously when $M_{\rm SUSY} = 10$ TeV. We found that such parameter space can be obtained when $\Delta m_h^2 \lesssim 0.48 M_{\rm SUSY}^2$, and then, we set $\Delta m_h^2 = 0.48 M_{\rm SUSY}^2$. % \footnote{To be more specific, the large $\Delta m_h^2$ makes the EWSB constraints severe since it gives large $| m_{h_u}^2 |$. As we can understand from Fig.~\ref{fig:parasp10}, large $r_h$ can evade the EWSB constraints. In such a case, however, small $\Delta y_b$ as well as small $s_{13}^{d L}$ are predicted.}, and larger $\Delta m_h^2$ predicts larger $s_{13}^{\psi L, R}$
On the other hand, we can obtain the parameter space in $M_{\rm SUSY} = 2, 100$ TeV case even when $\Delta m_h^2 = M_{\rm SUSY}^2$ since one of the EWSB and SUSY FCNC constraints is irrelevant in these cases.

From these figures, we can know which constraints determine the maximal value of $s_{13}^{d L}$. For $M_{\rm SUSY} = \mathcal{O}(1)$ TeV case, the SUSY FCNC constraints become the severest one and determine $s_{13}^{d L}$. Note that the EWSB constraints in this case can be satisfied in a wide area of parameter space. 
In contrast, the SUSY FCNC constraints become weaker and are irrelevant to the significant area for $M_{\rm SUSY} \gtrsim \mathcal{O}(100)$ TeV case. However, the EWSB constraints become severe because of the large values of $| m_{h_u}^2 |$ and $| m_{h_d}^2 |$. This means that a strong cancellation between $| m_{h_u}^2 |$ and $m_{h_d}^2$ is needed, and therefore, the allowed parameter space is constrained by the EWSB constraints. Interestingly, these constraints determine the size of $s_{13}^{d L}\simeq 9.63 \times 10^{-3}$ in our setup when the required SUSY threshold correction is obtained. 
In the middle $M_{\rm SUSY}$ case, like $M_{\rm SUSY} = 10$ TeV, both constraints relate to the maximal value of $s_{13}^{d L}$, as shown in Fig.~\ref{fig:parasp10}. 

It is notable that in our setup, the obtained mixing angles through the SUSY threshold corrections are related with each other due to the $SO(10)$ unification. This means that some of the predictions of FV process can be calculated once one of the predictions of the process are given. Therefore, we may be able to obtain the specific predictions and can test the models by the experiments.

Finally, we comment on the SM fermion masses. Since we do not consider the SUSY threshold corrections for $y_t$ and $y_{\tau}$ when we obtain the required SUSY threshold correction for $y_b$ and $\tan \beta$, the predictions of $y_t$ and $y_{\tau}$ are expected to slightly deviate from the experimental values. We found that the deviation of the top Yukawa coupling is $\mathcal{O}(1)$ \% or less, while that of $\tau$ Yukawa coupling is about $3 \sim 4$ \%. Note that the deviation of bottom Yukawa coupling should be small since we find the parameters for realizing the required SUSY threshold correction. The result for its deviation is less than $0.1$ \%, as expected.

%Note that the Yukawa couplings of first and second generations fermions are also deviate from each experimental value, and these deviations are larger than those of third generation fermions. 

\section{Conclusion}
\label{sec:conclusion}
In this paper, we considered the model with the Yukawa unification inspired by the $SO(10)$ GUT. In this model, we investigated the maximal size of the mixing angles of diagonalizing matrices for the Yukawa matrices, originated from the off-diagonal parts of SUSY threshold corrections. We need large off-diagonal parts of A-terms to obtain the sizable SUSY threshold corrections. However, these parts are constrained by the SUSY FCNC processes. Moreover, the CCB constraints also relate to the maximal value of A-terms. Therefore, we can estimate the maximal contributions to the mixing angles from the SUSY threshold corrections by using these constraints. 

We firstly found the required SUSY threshold correction and $\tan \beta$ for realizing the Yukawa unification. As reported in the other papers, the large $\tan \beta \sim 50$ and enough size of $\Delta y_b$ are needed. We emphasize that required size of $\Delta y_b$ is smaller than the maximal one when $M_{\rm SUSY} \leq 100$ TeV. Therefore, we can obtain required $\Delta y_b$ by choosing the input parameters at the GUT scale. 

Next, we showed the results for the $s_{13}^{\psi L, R}$ because this element has a possibility to reach the size of corresponding element of the CKM matrix, $s_{13}^{\rm CKM}$. As a result, we found that $s_{13}^{q L, R}$ $(q = u, d)$ can be comparable to $s_{13}^{\rm CKM}$ when $M_{\rm SUSY} > 10$ TeV. If the SUSY scale is $\mathcal{O}(1)$ TeV, the off-diagonal parts of A-terms cannot be large because of the SUSY FCNC constraints. Therefore, the results for $s_{13}^{q L, R}$ are about one order of magnitude smaller than $s_{13}^{\rm CKM}$. On the other hand, $M_{\rm SUSY} = 100$ TeV case is free from such constraints and predicts large $s_{13}^{q L, R}$, although the EWSB constraints become severe. For $M_{\rm SUSY} = \mathcal{O}(10)$ TeV case, both constraints are relevant to the maximal values of $s_{13}^{q L, R}$, and therefore, we have to find the appropriate value of $\Delta m_h^2$ to satisfy these constraints.

Importantly, the mixing angles obtained from the SUSY threshold corrections are related with each other when one considers the $SO(10)$ unification as in our setup. Therefore, if one of the FV processes is confirmed, the contributions to the other processes can be calculated, and we can test the models by the (future) experiments. In addition, if some experimental results of the FV process can be explained by these contributions, we may be able to obtain the signature of the $SO(10)$ unification. 
Furthermore, if the CKM matrix is realized by the other contributions in GUT models, we may find the contribution of the unification according to our results, although it deviates by the RGE effects.

\section*{Acknowledgments}
We thank B. Bajc for advise for starting this project and on present situation of Yukawa unification in SUSY SO(10) GUT. 
We also thank J. Hisano for fruitful discussion. 
Y.M. is supported in part by 
the National Natural Science Foundation of China (NNSFC) under 
contract Nos. 11435003, 11225523, and 11521064.
The work of Y.S. is supported by the Japan Society for the Promotion of Science (JSPS) Research Fellowships for Young Scientists, No. 16J08299.

\appendix

\section{SUSY threshold corrections}
\label{appendix:SUSYthreshold}
To calculate the SUSY threshold corrections, chirality-flipping self-energies $\Sigma^\psi_{LR}$ are important, 
which are calculated in Ref.~\cite{Crivellin:2011jt} in the decoupling limit $M_{\text{SUSY}}\gg v$.
We can divide %chirality-flipping self-energies 
$\Sigma^\psi_{LR}$ for $\psi=u,\, d,\, e$ as
\begin{eqnarray}
\Sigma^u_{LR} &=& \Sigma^{u\tilde{g}}_{LR}+\Sigma^{u\tilde{\chi}^0}_{LR}, \label{eq:uSigma} \\
\Sigma^d_{LR} &=& \Sigma^{d\tilde{g}}_{LR}+\Sigma^{d\tilde{\chi}^0}_{LR}+
\Sigma^{d\tilde{\chi}^\pm}_{LR}, \label{eq:dSigma} \\
\Sigma^e_{LR} &=& \Sigma^{e\tilde{\chi}^0}_{LR}+\Sigma^{e\tilde{\chi}^\pm}_{LR}. 
\label{eq:eSigma}
\end{eqnarray}
Each part of Eqs.~\eqref{eq:uSigma}, \eqref{eq:dSigma}, and \eqref{eq:eSigma} can be calculated by considering loop contributions from sfermions and fermionic partners of gauge and Higgs bosons. 
Therefore, we need to diagonalize the sfermion mass squared matrices, shown in Eqs.~\eqref{eq:up sfermion mass2}, \eqref{eq:down sfermion mass2}, and \eqref{eq:electron sfermion mass2}. 
%To calculate the SUSY threshold corrections we consider a sfermion loop, and therefore we diagonalize sfermion mass squared matrices.
%These sfermion mass squared matrices are shown in Eqs.~\eqref{eq:up sfermion mass2}, \eqref{eq:down sfermion mass2} and \eqref{eq:electron sfermion mass2}. %Eq.~\eqref{eq:up sfermion mass2}-\eqref{eq:electron sfermion mass2}.
For this calculation, we consider the effective Lagrangian within a decoupling limit $M_{SUSY}\gg v$, 
and hence, we neglect the chirality-flipping elements $\Delta^{\psi}_{LR}$ and $\Delta^{\psi}_{RL}$. %when we diagonalize each sfermion mass squared matrix.
Then the diagonalizing matrices for sfermion mass squared matrices, denoted as  $U_{\tilde{\psi}}$, can be defined by
\begin{equation}
U_{\tilde{\psi}}\hat{M}_{\tilde{\psi}}^2 U_{\tilde{\psi}}^{\dagger} = \hat{M}_{\tilde{\psi}\,diag}^2 = {\rm diag} \left( m_{\tilde{\psi}_{L1}}^2, m_{\tilde{\psi}_{L2}}^2, m_{\tilde{\psi}_{L3}}^2, m_{\tilde{\psi}_{R1}}^2, m_{\tilde{\psi}_{R2}}^2, m_{\tilde{\psi}_{R3}}^2 \right), ~
%\hat{M}_{\tilde{\psi}\,diag}^2,\,
U_{\tilde{\psi}} \equiv \left(
  \begin{array}{cc}
    U_{\tilde{\psi}}^L & 0 \\
    0 & U_{\tilde{\psi}}^R \\
  \end{array}
  \right).
\end{equation}
For convenience, we introduce
\begin{equation}
(\Lambda_m^{\psi \Gamma})_{ij}=( U_{\tilde{\psi}}^{\Gamma\,\dagger})_{im} ( U_{\tilde{\psi}}^{\Gamma})_{mj},
\end{equation}
where $\Gamma=L,\, R$ is the chirality. %index for chirality.
We summarize the chirality-flipping self-energies by using these parameters.

The chirality-flipping self-energies from gluino contribution are
\begin{align}
(\Sigma_{LR\,(mn)}^{d\tilde{g}})_{il} &= \frac{2\alpha_s}{3\pi} (-M_3) (\Lambda^{dL}_m)_{ij}(\Delta^d_{LR})_{jk}(\Lambda^{dR}_n)_{kl} C_0(|M_3|^2,\,m_{\tilde{d}_{Lm}}^2,\,m_{\tilde{d}_{Rn}}^2), \\[1.5ex]
%\end{align}
%\begin{align}
(\Sigma_{LR\,(mn)}^{u\tilde{g}})_{il} &= \frac{2\alpha_s}{3\pi} (-M_3) (\Lambda^{uL}_m)_{ij} (\Delta^u_{LR})_{jk}(\Lambda^{uR}_n)_{kl} C_0(|M_3|^2,\,m_{\tilde{u}_{Lm}}^2,\,m_{\tilde{u}_{Rn}}^2),
\end{align}
\begin{equation}
(\Sigma_{LR}^{\psi \tilde{g}})_{ij}=\sum_{m,n=1}^3
(\Sigma_{LR\,(mn)}^{\psi \tilde{g}})_{ij},
\end{equation}
%where $m_{\tilde{\psi}_{\Gamma n}}^2$ is eigenvalue for part of sfermion mass matrices $\Delta_{\Gamma \Gamma}^\psi$. 
where $C_0(m_1^2,\,m_2^2,\,m_3^2)$ is the loop function defined as
\begin{equation}
C_0(m_1^2,\,m_2^2,\,m_3^2)=\frac{m_1^2 m_2^2 \,\text{ln}\frac{m_1^2}{m_2^2}+
m_2^2 m_3^2 \,\text{ln}\frac{m_2^2}{m_3^2}+
m_3^2 m_1^2 \,\text{ln}\frac{m_3^2}{m_1^2}}
{(m_1^2-m_2^2)(m_2^2-m_3^2)(m_3^2-m_1^2)}.
\end{equation}

The chirality-flipping self-energies from neutralino contribution are
\begin{align}
(\Sigma_{LR\,(mn)}^{e\tilde{\chi}^0})_{il} &= \frac{1}{16\pi^2} 4 g_1^2 (-M_1) (\Lambda_m^{eL})_{ij} (\Delta_{LR}^e)_{jk} (\Lambda_n^{eR})_{kl} C_0(|M_1|^2,\,m_{\tilde{e}_{Lm}}^2,\,m_{\tilde{e}_{Rn}}^2), \\[1.0ex]
(\Sigma_{LR\,(m)}^{e\tilde{\chi}^0})_{ij} &= \frac{1}{16\pi^2} \biggl[\biggr. \frac{1}{\sqrt{2} g_2} M_W \sin \beta (\hat{Y}_e)_{jj} (\Lambda_m^{eL})_{ij} \Bigl\{\Bigr. g_2^2 (-M_2) \mu C_0(|M_2|^2,\,|\mu |^2,\,m_{\tilde{e}_{Lm}}^2) \nonumber \\
&\hspace{16.0em} - 4 g_1^2 (-M_1) \mu C_0(|M_1|^2,\,|\mu |^2,\,m_{\tilde{e}_{Lm}}^2) \Bigl.\Bigr\} \nonumber \\
&\hspace{4.0em} + 4 g_1^2 \frac{\sqrt{2} M_W \sin \beta}{g_2} (-M_1) \mu (\hat{Y}_e)_{ii} (\Lambda_m^{eR})_{ij} C_0(|M_1|^2,\,|\mu |^2,\,m_{\tilde{e}_{Rm}}^2) \biggl.\biggr], \\[1.5ex]
%\end{align}
%\begin{align}
(\Sigma_{LR\,(mn)}^{d\tilde{\chi}^0})_{il} &= \frac{1}{16\pi^2} \left( - \frac{1}{9} \right) 4 g_1^2 (-M_1) (\Lambda_m^{dL})_{ij}(\Delta_{LR}^d)_{jk}(\Lambda_n^{dR})_{kl} C_0(|M_1|^2,\,m_{\tilde{d}_{Lm}}^2,\,m_{\tilde{d}_{Rn}}^2), \\[1.0ex]
(\Sigma_{LR\,(m)}^{d\tilde{\chi}^0})_{ij} &= \frac{1}{16\pi^2} \biggl[\biggr. \frac{1}{\sqrt{2} g_2} M_W \sin \beta (\hat{Y}_d)_{jj}(\Lambda_m^{dL})_{ij} \Bigl\{\Bigr. g_2^2 (-M_2) \mu C_0(|M_2|^2,\,|\mu |^2,\,m_{\tilde{d}_{Lm}}^2) \nonumber \\
&\hspace{16.0em} + \frac{1}{3} 4 g_1^2 (-M_1) \mu C_0(|M_1|^2,\,|\mu |^2,\,m_{\tilde{d}_{Lm}}^2) \Bigl.\Bigr\} \nonumber \\
&\hspace{4.0em} + \frac{1}{3} 4 g_1^2 \frac{\sqrt{2} M_W \sin \beta}{g_2} (-M_1) \mu (\hat{Y}_d)_{ii} (\Lambda_m^{dR})_{ij}C_0(|M_1|^2,\,|\mu |^2,\,m_{\tilde{d}_{Rm}}^2) \biggl.\biggr], \\[1.5ex]
%\end{align}
%\begin{align}
(\Sigma_{LR\,(mn)}^{u\tilde{\chi}^0})_{il} &= \frac{1}{16\pi^2} \frac{2}{9} 4 g_1^2 (-M_1) (\Lambda_m^{uL})_{ij} (\Delta_{LR}^u)_{jk}(\Lambda_n^{uR})_{kl} C_0(|M_1|^2,\,m_{\tilde{u}_{Lm}}^2,\,m_{\tilde{u}_{Rn}}^2), \\[1.0ex]
(\Sigma_{LR\,(m)}^{u\tilde{\chi}^0})_{ij} &= 0,
\end{align}
\begin{equation}
(\Sigma_{LR}^{\psi \tilde{\chi}^0})_{ij}= \sum_{m,n=1}^3(\Sigma_{LR\,(mn)}^{\psi \tilde{\chi}^0})_{ij} + \sum_{m=1}^3(\Sigma_{LR\,(m)}^{\psi \tilde{\chi}^0})_{ij},
\end{equation}
where $M_W$ is the W boson mass and $g_i$ is the gauge coupling with GUT notation.

The chirality-flipping self-energies from chargino contribution are
\begin{align}
(\Sigma_{LR\,(mn)}^{d\tilde{\chi}^\pm})_{ij} &= - \frac{(\hat{Y}_d)_{jj}}{16\pi^2}\mu \Bigl\{\Bigl. (V_{CKM}^{\,\dagger})_{i3} (\Lambda_m^{dL})_{33} (V_{CKM})_{33}(\Delta_{LR}^{u\,*})_{33} (\Lambda_n^{uR})_{33} (\hat{Y}_u)_{33}\delta_{3j} \nonumber \\
&\hspace{20.0em} \times C_0(|\mu|^2,\,m_{\tilde{u}_{Lm}}^2,\,m_{\tilde{u}_{Rn}}^2) \Bigr.\Bigr\}, \\[1.0ex]
%\end{align}
%\begin{align}
(\Sigma_{LR\,(m)}^{d\tilde{\chi}^\pm})_{ij} &= \frac{(\hat{Y}_d)_{jj}}{16\pi^2}\mu \Bigl\{\Bigr. \sqrt{2} g_2 M_W \sin \beta (-M_2) (\Lambda_m^{dL})_{ij} C_0(m_{\tilde{d}_{Lm}}^2,\,|\mu|^2,\,|M_2|^2) \Bigl.\Bigr\}, \\[1.5ex]
%\end{align}
%\begin{align}
(\Sigma_{LR\,(mn)}^{e\tilde{\chi}^\pm})_{ij} &= 0, \\[1.0ex]
%\end{align}
%\begin{align}
(\Sigma_{LR\,(m)}^{e\tilde{\chi}^\pm})_{ij} &= \frac{(\hat{Y}_e)_{jj}}{16\pi^2}\mu \Bigl\{\Bigr. \sqrt{2} g_2 M_W \sin \beta (-M_2) (\Lambda_m^{eL})_{ij} C_0(m_{\tilde{e}_{Lm}}^2,\,|\mu|^2,\,|M_2|^2) \Bigl.\Bigr\},
\end{align}
\begin{equation}
(\Sigma_{LR}^{\psi \tilde{\chi}^\pm})_{ij} = \sum_{m,n=1}^3(\Sigma_{LR\,(mn)}^{\psi \tilde{\chi}^\pm})_{ij} + \sum_{m=1}^3(\Sigma_{LR\,(m)}^{\psi \tilde{\chi}^\pm})_{ij}.
\end{equation}
Note that we neglect the chargino contributions to up-quark self-energies $(\Sigma_{LR}^{u\tilde{\chi}^\pm})_{ij}$ since these contributions are not $\tan \beta$-enhanced ones.

%To calculate chirality-flipping self-energies we have to use loop function which is given by
%\begin{equation}
%C_0(m_1^2,\,m_2^2,\,m_3^2)=\frac{m_1^2 m_2^2 \,\text{ln}\frac{m_1^2}{m_2^2}+
%m_2^2 m_3^2 \,\text{ln}\frac{m_2^2}{m_3^2}+
%m_3^2 m_1^2 \,\text{ln}\frac{m_3^2}{m_1^2}}
%{(m_1^2-m_2^2)(m_2^2-m_3^2)(m_3^2-m_1^2)}.
%\end{equation}

%To calculate SM fermion mass matrices $(m_\psi^{(0)})_{ij}$ from SUSY fermion mass matrices we add chirality-flipping self-energies to SUSY fermion mass matrices $\hat{m}_{\psi\, i} \delta_{ij}$ as 
At the SUSY scale, the above corrections are added to the fermion mass matrices as
\begin{equation}
(m_\psi^{(0)})_{ij} \equiv \hat{m}_{\psi\, i} \delta_{ij} + (\Sigma_{LR}^\psi)_{ij}.
\end{equation}
Then, the chirality-flipping self-energies $\Sigma_{LR}^\psi$ and the SUSY threshold corrections $\Delta y_\psi$ defined in Eq.~(\ref{eq:matching}) have following relations:
\begin{equation}
\Delta y_t = - (\Sigma_{LR}^u)_{33}/(v \sin \beta),~~~ \Delta y_{b,\,\tau} = - (\Sigma_{LR}^{d,\,e})_{33}/(v \cos \beta).
\end{equation}

\section{Constraints on the SUSY contributions}
\label{appendix:ConstMIP}
%\section{Constraint for the SUSY threshold corrections}
%mass insertion parameter and color charge breaking
In this Appendix, we summarize the constraints on the SUSY contributions from the SUSY FCNC and CCB constraints. 
%, shown in Eq.~\eqref{eq:MIpara}. These parameters are constrained by the SUSY FCNC processes and the CCB bounds.
 % for the SUSY threshold corrections. 
%%% below equation and sentences around it should be moved to main text %%%
%First of all, the definition of MIPs used in this paper is
%\begin{eqnarray}
%  \left( \delta_{AB}^f \right)_{ij} \equiv \frac{\left( \Delta_{AB}^f \right)_{ij}}{\sqrt{\hat{M}_{\tilde{f} \, ii}^2 \hat{M}_{\tilde{f} \, jj}^2}},
%  \label{eq:MIpara}
%\end{eqnarray}
%where $A, B = L$ or $R$, $f = u, d, e$ and $i, j = 1, 2, 3$. $\left( \Delta_{AB}^f \right)_{ij}$ is matrix element of $6 \times 6$ sfermion mass matrix defined by Eqs. \eqref{eq:up sfermion mass2}, \eqref{eq:down sfermion mass2} and \eqref{eq:electron sfermion mass2}. This parameter is constrained by the SUSY FCNC processes and the CCB bounds.

First, we will show the SUSY FCNC constraints. For the squark sector, the upper bounds on SUSY contributions are mainly come from the constraints of the neutral meson mixing. This can be calculated by the general forms which can be found in Ref.~\cite{Gabbiani:1996hi}, using the MIPs. 
%\begin{align}
%\langle M^0 | \mathcal{H}_{\rm eff}^{\Delta F = 2} | \overline{M}^0 \rangle = - \frac{\alpha_s^2}{216 m_{\tilde{q}}^2} & \frac{1}{3} m_M f_M^2 \nonumber \\
%\times \Biggl\{&\Biggr. \left[ \left( \delta_{LL}^q \right)_{ij}^2 + \left( \delta_{RR}^q \right)_{ij}^2 \right] \left( 24 x f_6 (x) + 66 \tilde{f}_6 (x) \right) \nonumber \\
%& + \left( \delta_{LL}^q \right)_{ij} \left( \delta_{RR}^q \right)_{ij} \left[ \left( 384 \left( \frac{m_M}{m_{q_i} + m_{q_j}} \right)^2 + 72 \right) x f_6 (x) \right. \nonumber \\
%&\hspace{6.5em} + \left. \left( -24 \left( \frac{m_M}{m_{q_i} + m_{q_j}} \right)^2 + 36 \right) \tilde{f}_6 (x) \right] \nonumber \\
%& + \left[ \left( \delta_{LR}^q \right)_{ij}^2 + \left( \delta_{RL}^q \right)_{ij}^2 \right] \left( -132 \left( \frac{m_M}{m_{q_i} + m_{q_j}} \right)^2 \right) x f_6 (x) \nonumber \\
%& + \left( \delta_{LR}^q \right)_{ij} \left( \delta_{LR}^q \right)_{ij} \left( -144 \left( \frac{m_M}{m_{q_i} + m_{q_j}} \right)^2 - 84 \right) \tilde{f}_6 (x) \Biggl.\Biggr\},
%\label{eq:HeffMSSM}
%\end{align}
%where $m_M$ and $f_M$ are the mass and decay constant for the meson $M$, $x = m_{\tilde{g}}^2 / m_{\tilde{q}}^2$ with the gluino mass $m_{\tilde{g}}$ and average squark mass $m_{\tilde{q}}$, and $f_6 (x)$ and $\tilde{f}_6 (x)$ are loop functions defined as
%\begin{align}
%f_6 (x) &= \frac{6 (1 + 3 x) \ln x + x^3 - 9 x^2 - 9 x + 17}{6 (x-1)^5}, \\
%\tilde{f}_6 (x) &= \frac{6 x (1 + x) \ln x - x^3 - 9 x^2 + 9 x + 1}{3 (x-1)^5}.
%\end{align}
In our calculation, $B_d$-$\overline{B_d}$ mixing is important. In order to check the experimental bounds, it is useful to define the following variables:
\begin{align}
C_{B_d} e^{2 i \phi_{B_d}} \equiv \frac{\langle B_d^0 | \mathcal{H}_{\rm eff}^{\rm tot} | \overline{B_d}^0 \rangle}{\langle B_d^0 | \mathcal{H}_{\rm eff}^{\rm SM} | \overline{B_d}^0 \rangle} = 1 + \frac{\langle B_d^0 | \mathcal{H}_{\rm eff}^{\rm NP} | \overline{B_d}^0 \rangle}{\langle B_d^0 | \mathcal{H}_{\rm eff}^{\rm SM} | \overline{B_d}^0 \rangle},
\end{align}
where $\mathcal{H}_{\rm eff}^{\rm SM, NP}$ is the effective Hamiltonian of the SM and the MSSM, respectively. In this notation, $C_{B_d}$ and $\phi_{B_d}$ are related to the observables as
\begin{align}
\Delta M_{B_d} = C_{B_d} \Delta M_{B_d}^{\rm SM}, ~~~ \sin 2 \beta_d = \sin \left( 2 \beta_d^{\rm SM} + 2 \phi_{B_d} \right),
\end{align}
where $\Delta M_{B_d}$ is the mass difference of $B_d$ meson and $2 \beta_d$ is the phase of $\langle B_d^0 | \mathcal{H}_{\rm eff}^{\rm tot} | \overline{B_d}^0 \rangle$. For the bounds of $C_{B_d}$ and $\phi_{B_d}$, and the SM prediction of $\beta_d^{\rm SM}$, we use results of the UTfit collaboration \cite{UTfit}:\footnote{See also \href{http://www.utfit.org/UTfit/WebHome}{http://www.utfit.org/UTfit/WebHome} for the latest results.} $0.83 \leq C_{B_d} \leq 1.29$ and $-6.0^{\circ} \leq \phi_{B_d} \leq 1.5^{\circ}$ with $2\sigma$ level, and $\sin 2 \beta_d^{\rm SM} = 0.738$.

The upper bounds on the SUSY contributions in slepton sector are mainly come from the constraints of $l_i \to l_j \gamma$ process. The branching ratio of this process can be calculated by chargino and neutralino contributions as
\begin{equation}
  {\rm BR}(l_i \to l_j \gamma) = \frac{\alpha}{4} \left( \frac{m_{l_i}^5}{\Gamma_{l_i}} \right) \Bigl[ |(A^L_{ij})|^2 + |(A^R_{ij})|^2 \Bigr].
  \label{eq:litoljgamma}
\end{equation}
$(A^L_{ij})$ and $(A^R_{ij})$ can be estimated by the mass insertion approximation, and these expressions can be found in Ref.~\cite{Paradisi:2005fk}.
Note that the current experimental upper bounds of ${\rm BR}(l_i \to l_j \gamma)_{\rm exp}$ are summarized in Ref.~\cite{PDG}: ${\rm BR}(\mu \to e \gamma) < 4.2 \times 10^{-13}$, ${\rm BR}(\tau \to e \gamma) < 3.3 \times 10^{-8}$, and ${\rm BR}(\tau \to \mu \gamma) < 4.4 \times 10^{-8}$. 
We emphasize that the bound of $\tau \to e \gamma$ is satisfied in our calculation, even when $M_{\rm SUSY} = 2$ TeV.

%%%%%%%%%%%%%%%%%
%%% CCB bound %%%
%%%%%%%%%%%%%%%%%

In addition to above constraints, there are CCB constraints, which can be read as \cite{CCB}
\begin{align}
  |(\hat{A}_u)_{ij}|^2 &\leq (\hat{Y}_u)_{kk} \left( m_{\tilde{u}_{L i}}^2 + m_{\tilde{u}_{R j}}^2 + m_{h_u}^2 \right), \label{eq:AuCCB} \\[0.5ex]
  |(\hat{A}_d)_{ij}|^2 &\leq (\hat{Y}_d)_{kk} \left( m_{\tilde{d}_{L i}}^2 + m_{\tilde{d}_{R j}}^2 + m_{h_d}^2 \right), \label{eq:AdCCB} \\[0.5ex]
  |(\hat{A}_e)_{ij}|^2 &\leq (\hat{Y}_e)_{kk} \left( m_{\tilde{e}_{L i}}^2 + m_{\tilde{e}_{R j}}^2 + m_{h_d}^2 \right), \label{eq:AeCCB}
\end{align}
where $k = \max(i,j)$. 
%%%
Note that because of the sfermion mass spectrum, the bound for $(\hat{A}_{\psi})_{33}$ is stronger than the other ones.
%%%

%\section{Required and maximal SUSY threshold correction}
\section{Maximal SUSY threshold correction}
\label{appendix:RM_SUSYthreshold}

In Fig.~\ref{fig:req_SUSYthreshold}, we showed the required SUSY threshold correction to realize $SO(10)$ Yukawa unification.
Another important issue is whether the maximal SUSY threshold correction can be larger than the required one.
The SUSY threshold corrections for the bottom Yukawa coupling mainly come from the gluino contribution:
\begin{equation}
\Delta y_b \simeq -\frac{2}{3}\frac{\alpha_s}{\pi} M_3(- a_b + \mu y_b \tan \beta ) C_0 (|M_3|^2, m_{\tilde{b}_L}^2, m_{\tilde{b}_R}^2),
\label{eq:Deltayb}
\end{equation}
where $m_{\tilde{b}_L}^2 = m_{\tilde{d}_{L 3}}^2$ and $m_{\tilde{b}_R}^2 = m_{\tilde{d}_{R 3}}^2$. Similar notations for $m_{\tilde{t}_{L, R}}^2$ and $m_{\tilde{\tau}_{L, R}}^2$ are used.
%$C_0 (m_1^2, m_2^2, m_3^2)$ is the loop function which is defined in Eq.~(\ref{eq:loopF_def}).
In this paper, the SUSY scale is defined as 
\begin{equation}
M_{\rm SUSY} = \sqrt{m_{\tilde{t}_L}(M_{\text{SUSY}}) m_{\tilde{t}_R}(M_{\text{SUSY}})}
\end{equation}
and the gluino mass $M_3 = M_{\rm SUSY}$ at $M_{\rm SUSY}$.
%, and therefore, the sbottom masses $m_{\tilde{b}_L}^2$, $m_{\tilde{b}_R}^2$ are fixed when we use the constrained MSSM assumption.
%For the discussion about the maximal SUSY threshold correction, 
Hereafter, we also assume $m_{\tilde{b}_L}^2 = m_{\tilde{t}_L}^2$ and $m_{\tilde{t}_L}^2 + m_{\tilde{t}_R}^2 = 2 M_{\rm SUSY}^2$ for simplicity. 
%Once we fix the remaining parameters, $a_b$ and $\mu$, we can estimate $\Delta y_b$.
As one can know from Eq.~\eqref{eq:Deltayb}, the large negative $a_b$ and large positive $\mu$ are important to maximize $| \Delta y_b |$.

However, $| a_b |$ and $\mu$ have upper limits which come from the CCB and EWSB constraints.
%The CCB constraints are discussed at the Appendix \ref{appendix:ConstMIP}.
The maximal value of $| a_b |$ can be estimated as $\sqrt{y_b (m_{\tilde{b}_L}^2 + m_{\tilde{b}_R}^2 + m_{h_d}^2)}$, according to Eq.~\eqref{eq:AdCCB}.
On the other hand, that of $\mu$ is complicated because $\mu$ depends on the MSSM Higgs mass parameters, $m_{h_u}^2$ and $m_{h_d}^2$, through the tree-level condition to have the EWSB minimum in the Higgs potential \cite{Inoue:1982ej}
%The EWSB constraints are following tree-level condtion to have the EWSB minimum in the Higgs potential \cite{Inoue:1982ej} 
\begin{equation}
| \mu |^2 = \frac{m_{h_d}^2 - m_{h_u}^2 \tan^2 \beta}{\tan^2 \beta - 1} - \frac{m_Z^2}{2}. %\simeq - m_{h_u}^2,
\label{eq:mu_def}
\end{equation}
%where the last approximation is the case with large $\tan \beta$. 
Note that the EWSB can be occurred when the following conditions are satisfied:
\begin{equation}
m_{h_u}^2 + | \mu |^2 < 0, ~~~ m_{h_d}^2 + | \mu |^2 > 0, ~~~ \text{and} ~~~ 2 | \mu |^2 + m_{h_u}^2 + m_{h_d}^2 > 0.
\label{eq:EWSB_const}
\end{equation}
Therefore, in the following discussion, we assume the case with $m_{h_u}^2 < 0$ to satisfy the first inequality of Eq.~\eqref{eq:EWSB_const}.
%When we consider these constraints we can calculate the maximal SUSY threshold corrections.

From Eq.~\eqref{eq:Deltayb}, $\Delta y_b$ is dominated by the contribution from $\mu$ because of the large $\tan \beta$ enhancement. Therefore, we first discuss the maximal value of $\mu$ which satisfies the CCB and EWSB constraints. It is clear that $m_{h_u}^2$ is important to obtain the maximal value of $\mu$. %, so that we discuss the maximal value of $m_{h_u}^2$ in the following two cases: $m_{h_d}^2 > 0$ and $m_{h_d}^2 < 0$. 
Note that the effect of $a_b$ is discussed later.

The maximal value of $| m_{h_u}^2 |$ is $m_{\tilde{t}_L}^2 + m_{\tilde{t}_R}^2 = 2 M_{\rm SUSY}^2$ from the CCB constraint of $| ( \hat{A}_u )_{33} |$. Although the large $m_{h_d}^2$ is needed to maximize $\mu$, the EWSB constraints cannot be satisfied with $| m_{h_u}^2 | = 2 M_{\rm SUSY}^2$ in the large $M_{\rm SUSY}$ when $m_{h_d}^2 > 0$. This can be understood from Eqs.~\eqref{eq:mu_def} and \eqref{eq:EWSB_const}. By combining these equation and inequalities, we can obtain following inequalities:
\begin{align}
%\frac{m_Z^2}{2} \frac{\tan^2 \beta - 1}{\tan^2 \beta} < | m_{h_u}^2 | + m_{h_d}^2 < \frac{m_Z^2}{2} \left( \tan^2 \beta - 1 \right).
\frac{\tan^2 \beta - 1}{\tan^2 \beta + 1} \: m_Z^2 < | m_{h_u}^2 | + m_{h_d}^2 < \frac{m_Z^2}{2} \left( \tan^2 \beta - 1 \right).
\label{eq:EWSBneed}
\end{align}
Then, a large $| m_{h_u}^2 |$ needs a small $m_{h_d}^2$ in order to satisfy the right part of the inequalities. This means that $m_{h_d}^2$ becomes negative when $M_{\rm SUSY} \gtrsim M_{\rm th}$ where $M_{\rm th} \equiv \frac{m_Z}{2} \sqrt{\tan^2 \beta - 1} \approx 2.25$ TeV when $\tan \beta \approx 50$. 
%This means that the maximal values of $| m_{h_u}^2 |$ and $m_{h_d}^2$ with $m_{h_d}^2 > 0$ is
%\begin{align}
%| m_{h_u}^2 | &= 
%\begin{cases}
%2 M_{\rm SUSY}^2 & \text{when $M_{\rm SUSY} \lesssim M_{\rm th}$,} \\
%\frac{m_Z^2}{2} \left( \tan^2 \beta - 1 \right) & \text{when $M_{\rm SUSY} \gtrsim M_{\rm th}$,}
%\end{cases} \\
%m_{h_d}^2 &= \frac{m_Z^2}{2} \left( \tan^2 \beta - 1 \right) - | m_{h_u}^2 |,
%\end{align}
%When $m_{h_d}^2 < 0$, on the other hand, Eq.~\eqref{eq:EWSBneed} can be always satisfied with $| m_{h_u}^2 | = 2 M_{\rm SUSY}^2$ by adjusting $m_{h_d}^2$ appropriately. Therefore, in order to maximize $\mu$ in this case,
%\begin{align}
%| m_{h_u}^2 | &= 2 M_{\rm SUSY}^2, \\
%m_{h_d}^2 &=
%\begin{cases}
%0 & \text{when $M_{\rm SUSY} \lesssim M_{\rm th}$,} \\
%\frac{m_Z^2}{2} \left( \tan^2 \beta - 1 \right) - 2 M_{\rm SUSY}^2 & \text{when $M_{\rm SUSY} \gtrsim M_{\rm th}$.}
%\end{cases}
%\end{align}
As a result, the values of $| m_{h_u}^2 |$ and $m_{h_d}^2$ which maximize $\mu$ are
\begin{align}
| m_{h_u}^2 | = 2 M_{\rm SUSY}^2, ~~~ m_{h_d}^2 = \frac{m_Z^2}{2} \left( \tan^2 \beta - 1 \right) - 2 M_{\rm SUSY}^2,
\end{align}
and the maximal $\mu$ can be obtained from Eq.~\eqref{eq:mu_def} as
%Compared with these two cases, we can obtain the maximal value of $\mu$ with $m_{h_d}^2 > 0$ when $M_{\rm SUSY} \lesssim M_{\rm th}$ and with $m_{h_d}^2 < 0$ when $M_{\rm SUSY} \gtrsim M_{\rm th}$, whose size is
\begin{align}
\mu_{\rm max} = \sqrt{| m_{h_u}^2 |} = \sqrt{2} M_{\rm SUSY}.
\label{eq:maxmu}
\end{align}

In this case, the CCB constraint of $\hat{A}_u$ is $| (\hat{A_u})_{ij} |^2 \leq 0$, which leads to $| ( \hat{A}_u )_{33} | = | a_t | = 0$. 
%Due to the Yukawa unification and the assumption for A-terms at the GUT scale, $a_b$ should be zero. 
However, that of $| a_b |$ is
\begin{align}
| a_b |^2 &\leq (\hat{Y}_b)_{33} \left[ m^2_{\tilde{b}_L} + m^2_{\tilde{b}_R} + \frac{m_Z^2}{2} \left( \tan^2 \beta - 1 \right) - 2 M_{\rm SUSY}^2 \right] \nonumber \\
&= y_b \left[ \frac{m_Z^2}{2} \left( \tan^2 \beta - 1 \right) + \delta m^2_{bt} \right],
\label{eq:maxab}
\end{align}
where $\delta m^2_{bt} \equiv m^2_{\tilde{b}_L} + m^2_{\tilde{b}_R} - 2 M_{\rm SUSY}^2 = m^2_{\tilde{b}_R} - m^2_{\tilde{t}_R}$. Note that $\delta m^2_{bt}$ is positive since $m^2_{\tilde{b}_R} > m^2_{\tilde{t}_R}$ is realized at the SUSY scale through the MSSM RGE. We found that $\delta m^2_{bt} \sim 0.08 M_{\rm SUSY}^2$ in our calculation. Therefore, the maximal value of $| a_b |$ is
\begin{align}
| a_b |_{\rm max} = \sqrt{y_b \left[ \frac{m_Z^2}{2} \left( \tan^2 \beta - 1 \right) + \delta m^2_{bt} \right]}.
\label{eq:abmax}
\end{align}
%in all value of $M_{\rm SUSY}$.

We emphasize that Eq.~\eqref{eq:abmax} is obtained by assuming that the contribution from $\mu$ to $\Delta y_b$ is dominated. In order to check this and to see the dependence of $a_b$, we also consider the case where $| a_b |_{\rm max}$ is the following expression:
\begin{align}
| a_b |_{\rm max} = \sqrt{y_b \left[ \frac{m_Z^2}{2} \left( \tan^2 \beta - 1 \right) + \delta m^2_{bt} + M^2 \right]}.
\label{eq:abmaxM}
\end{align}
Here, $M^2 \geq 0$ is a parameter for taking $| a_b |_{\rm max}$ to be large. This modification can be justified by changing $m_{h_u}^2$ and $m_{h_d}^2$ as
\begin{align}
m_{h_u}^2 = - 2 M_{\rm SUSY}^2 + M^2, ~~~ m_{h_d}^2 = \frac{m_Z^2}{2} \left( \tan^2 \beta - 1 \right) - 2 M_{\rm SUSY}^2 + M^2
\end{align}
without conflicting with the CCB and EWSB constraints. Note that since $m_{h_u}^2 < 0$ is needed for the correct EWSB, the range for $M^2$ is $0 \leq M^2 < 2 M_{\rm SUSY}^2$. This modification affects the value of $\mu_{\rm max}$ through $m_{h_u}^2$: $\mu_{\rm max} = \sqrt{| m_{h_u}^2 |} = \sqrt{2 M_{\rm SUSY}^2 - M^2}$.

\begin{figure}[tb]
  \begin{center}
  \includegraphics[width=0.6\textwidth,bb= 0 0 450 292]{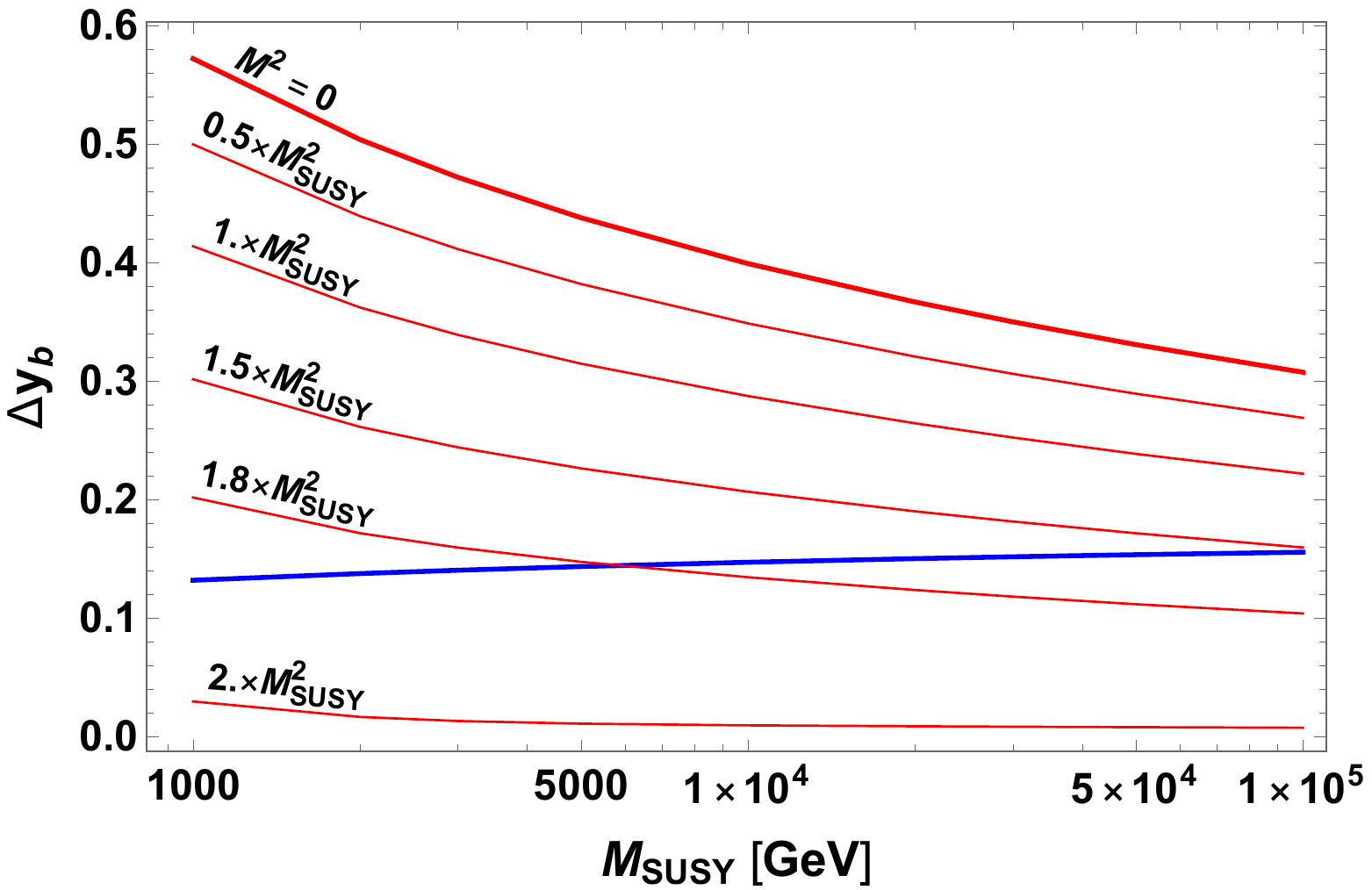}
  \caption{The $M_{\rm SUSY}$ dependence of the required and maximal SUSY threshold corrections. The blue line shows the required SUSY threshold correction for realizing the $SO(10)$ Yukawa unification. The red lines show the maximal SUSY threshold corrections with different values of $M^2$. The thick red line which is the case with $M^2 = 0$ corresponds to the maximal SUSY threshold correction in the range of $0 \leq M^2 < 2 M_{\rm SUSY}^2$.} %For this figure, we assume $\delta m^2_{bt} = 0.08 M_{\rm SUSY}^2$}
  \label{fig:maxDelyb}
  \end{center}
\end{figure}

By using above expressions for $m_{h_u}^2$, $m_{h_d}^2$, and $a_b$, we can calculate the maximal value of $\Delta y_b$ in each $M_{\rm SUSY}$. We show the results in Fig.~\ref{fig:maxDelyb}.
The blue line shows the required SUSY threshold correction for realizing the $SO(10)$ Yukawa unification, while the red ones show the maximal SUSY threshold corrections with different values of $M^2$, calculated from the above discussion. The thick red line which is the case with $M^2 = 0$ corresponds to the maximal SUSY threshold correction in the range of $0 \leq M^2 < 2 M_{\rm SUSY}^2$. Here, we assume that $M_3 = M_{\rm SUSY}$ as in our setup, $C_0 (|M_3|^2, m_{\tilde{b}_L}^2, m_{\tilde{b}_R}^2) = - \frac{1}{2 M_{\rm SUSY}^2}$, and $\delta m^2_{bt} = 0.08 M_{\rm SUSY}^2$. The second assumption is valid when $m_{\tilde{b}_{L,R}}^2 \simeq M_{\rm SUSY}^2$. Note that $\delta m^2_{bt}$ is not important to obtain the maximal value of $\Delta y_b$ unless this is enough large to compete the other contributions. We found that $\delta m^2_{bt}$ is less than $0.1 M_{\rm SUSY}^2$ in our calculation since we use the universal sfermion mass as the input at the GUT scale.

We can find important results from Fig.~\ref{fig:maxDelyb}. First, the maximal SUSY threshold correction is larger than the required one when $M_{\rm SUSY} \leq 100$ TeV, and then, we can find the parameter set which realizes the required SUSY threshold correction. Second, the maximal SUSY threshold correction decreases when $M^2$ becomes large. This means that $\Delta y_b$ is dominated by the contribution from $\mu$, as one can understand from the plot with $M^2 = 2 M_{\rm SUSY}^2$ case, namely, $\mu_{\rm max} = 0$ case. Therefore, we can obtain the required SUSY threshold correction from the appropriate values of $m_{h_u}^2$ and $m_{h_d}^2$. Third, as the SUSY scale increases, the required SUSY threshold correction increases, while the maximal SUSY threshold correction decreases. This is because the rough expression for the maximal value of $\Delta y_b$ is 
\begin{align}
\Delta y_b \sim - \frac{2}{3} \frac{\alpha_s}{\pi} M_{\rm SUSY} \bigl( \sqrt{2} M_{\rm SUSY} y_b \tan \beta \bigr) \left( - \frac{1}{2 M_{\rm SUSY}^2} \right) \propto \alpha_s y_b \tan \beta,
\end{align}
and this combination becomes small when the SUSY scale become high. Therefore, we can expect that there is an intersection point between the required and maximal SUSY threshold corrections. This means that there is an upper bound on the SUSY scale when we consider the $SO(10)$ Yukawa unification, although its bound is expected to be far above $M_{\rm SUSY} = 100$ TeV. 

Finally, we comment on the situation in which we do not consider in the above discussion. If $m_{\tilde{\tau}_L}^2 + m_{\tilde{\tau}_R}^2 < 2 M_{\rm SUSY}^2 - \frac{m_Z^2}{2} \left( \tan^2 \beta - 1 \right)$ due to the small $m_{\tilde{\tau}_R}^2$ through the MSSM RGE, the maximal value of $| m_{h_d}^2 |$ is $m_{\tilde{\tau}_L}^2 + m_{\tilde{\tau}_R}^2$ from the CCB constraint of $( \hat{A}_e )_{33}$ when $m_{h_d}^2 < 0$. In this case, the maximal value of $| m_{h_u}^2 |$ is modified as
\begin{align}
| m_{h_u}^2 | = m_{\tilde{\tau}_L}^2 + m_{\tilde{\tau}_R}^2 + \frac{m_Z^2}{2} \left( \tan^2 \beta - 1 \right)
\end{align}
in order to satisfy the EWSB constraints. This is smaller than $2 M_{\rm SUSY}^2$, and therefore, the maximal value of $\mu$ is also smaller than the above case. Note that in this case, since $| a_b |_{\rm max} = \sqrt{y_b \, \delta m^2_{b \tau}}$ where $\delta m^2_{b \tau} \equiv m_{\tilde{b}_L}^2 + m_{\tilde{b}_R}^2 - \left( m_{\tilde{\tau}_L}^2 + m_{\tilde{\tau}_R}^2 \right) < 2 M_{\rm SUSY}^2$, we can safely ignore the contribution from $a_b$, according to the above results. Therefore, we can obtain the maximal $\Delta y_b$ from Fig.~\ref{fig:maxDelyb} straightforwardly. For example, when $| m_{h_u}^2 | \simeq 1.5 M_{\rm SUSY}^2$, predicted value of $\Delta y_b$ is around the red line with $M^2 = 0.5 M_{\rm SUSY}^2$. However, %as mentioned above, 
$m_{\tilde{\tau}_L}^2 + m_{\tilde{\tau}_R}^2$ is not far below $2 M_{\rm SUSY}^2$ because of the input parameter at the GUT scale. Therefore, we can conclude that the maximal SUSY threshold correction is always larger than the required one when $M_{\rm SUSY} \leq 100$ TeV.


\begin{thebibliography}{99}
\bibitem{Aad:2012tfa} 
  G.~Aad {\it et al.} [ATLAS Collaboration],
  %``Observation of a new particle in the search for the Standard Model Higgs boson with the ATLAS detector at the LHC,''
  \href{https://doi.org/10.1016/j.physletb.2012.08.020}{Phys.\ Lett.\ B {\bf 716}, 1 (2012)}
  %doi:10.1016/j.physletb.2012.08.020
  [\href{https://arxiv.org/abs/1207.7214}{arXiv:1207.7214 [hep-ex]}];
  %%CITATION = doi:10.1016/j.physletb.2012.08.020;%%
  %9303 citations counted in INSPIRE as of 11 Mar 2019
  S.~Chatrchyan {\it et al.} [CMS Collaboration],
  %``Observation of a new boson at a mass of 125 GeV with the CMS experiment at the LHC,''
  \href{https://doi.org/10.1016/j.physletb.2012.08.021}{Phys.\ Lett.\ B {\bf 716}, 30 (2012)}
  %doi:10.1016/j.physletb.2012.08.021
  [\href{https://arxiv.org/abs/1207.7235}{arXiv:1207.7235 [hep-ex]}].
  %%CITATION = doi:10.1016/j.physletb.2012.08.021;%%
  %9088 citations counted in INSPIRE as of 11 Mar 2019
  
\bibitem{CKMmatrix} 
  N.~Cabibbo,
  %``Unitary Symmetry and Leptonic Decays,''
  \href{https://doi.org/10.1103/PhysRevLett.10.531}{Phys.\ Rev.\ Lett.\  {\bf 10}, 531 (1963)};
  %doi:10.1103/PhysRevLett.10.531
  %%CITATION = doi:10.1103/PhysRevLett.10.531;%%
  %5592 citations counted in INSPIRE as of 02 Oct 2017
  M.~Kobayashi and T.~Maskawa,
  %``CP Violation in the Renormalizable Theory of Weak Interaction,''
  \href{https://doi.org/10.1143/PTP.49.652}{Prog.\ Theor.\ Phys.\  {\bf 49}, 652 (1973)}.
  %doi:10.1143/PTP.49.652
  %%CITATION = doi:10.1143/PTP.49.652;%%
  %9219 citations counted in INSPIRE as of 02 Oct 2017

\bibitem{Hall:1985dx} 
  L.~J.~Hall, V.~A.~Kostelecky and S.~Raby,
  %``New Flavor Violations in Supergravity Models,''
  \href{https://doi.org/10.1016/0550-3213(86)90397-4}{Nucl.\ Phys.\ B {\bf 267}, 415 (1986)}.
  %doi:10.1016/0550-3213(86)90397-4
  %%CITATION = doi:10.1016/0550-3213(86)90397-4;%%
  %743 citations counted in INSPIRE as of 02 Oct 2017

\bibitem{Georgi:1974sy} 
  H.~Georgi and S.~L.~Glashow,
  %``Unity of All Elementary Particle Forces,''
  \href{https://doi.org/10.1103/PhysRevLett.32.438}{Phys.\ Rev.\ Lett.\  {\bf 32}, 438 (1974)}.
  %doi:10.1103/PhysRevLett.32.438
  %%CITATION = doi:10.1103/PhysRevLett.32.438;%%
  %4428 citations counted in INSPIRE as of 02 Oct 2017

\bibitem{SKFVPC}
  K.~Abe {\it et al.} [Super-Kamiokande Collaboration],
  %``Search for proton decay via $p \to e^+\pi^0$ and $p \to \mu^+\pi^0$ in 0.31\UTF{2009}\UTF{2009}megaton\UTF{00B7}years exposure of the Super-Kamiokande water Cherenkov detector,''
  \href{https://doi.org/10.1103/PhysRevD.95.012004}{Phys.\ Rev.\ D {\bf 95}, 012004 (2017)}
  %doi:10.1103/PhysRevD.95.012004
  [\href{https://arxiv.org/abs/1610.03597}{arXiv:1610.03597 [hep-ex]}];
  %%CITATION = doi:10.1103/PhysRevD.95.012004;%%
  %20 citations counted in INSPIRE as of 02 Oct 2017
  K.~Abe {\it et al.} [Super-Kamiokande Collaboration],
  %``Search for nucleon decay into charged antilepton plus meson in 0.316 megaton$\cdot$years exposure of the Super-Kamiokande water Cherenkov detector,''
  \href{https://doi.org/10.1103/PhysRevD.96.012003}{Phys.\ Rev.\ D {\bf 96}, 012003 (2017)}
  %doi:10.1103/PhysRevD.96.012003
  [\href{https://arxiv.org/abs/1705.07221}{arXiv:1705.07221 [hep-ex]}].
  %%CITATION = doi:10.1103/PhysRevD.96.012003;%%
  %2 citations counted in INSPIRE as of 02 Oct 2017

\bibitem{FVPD}
  Y.~Achiman and C.~Merten,
  %``Large quark rotations, neutrino mixing and proton decay,''
  \href{https://doi.org/10.1016/S0550-3213(00)00353-9}{Nucl.\ Phys.\ B {\bf 584}, 46 (2000)}
  %doi:10.1016/S0550-3213(00)00353-9
  [\href{https://arxiv.org/abs/hep-ph/0004023}{hep-ph/0004023}];
  %%CITATION = doi:10.1016/S0550-3213(00)00353-9;%%
  %8 citations counted in INSPIRE as of 02 Oct 2017
  Y.~Achiman and M.~Richter,
  %``Gauge mediated proton decay in a renormalizable SUSY SO(10) with realistic mass matrices,''
  \href{https://doi.org/10.1016/S0370-2693(01)01357-0}{Phys.\ Lett.\ B {\bf 523}, 304 (2001)}
  %doi:10.1016/S0370-2693(01)01357-0
  [\href{https://arxiv.org/abs/hep-ph/0107055}{hep-ph/0107055}];
  %%CITATION = doi:10.1016/S0370-2693(01)01357-0;%%
  %10 citations counted in INSPIRE as of 02 Oct 2017
  J.~R.~Ellis, D.~V.~Nanopoulos and J.~Walker,
  %``Flipping SU(5) out of trouble,''
  \href{https://doi.org/10.1016/S0370-2693(02)02956-8}{Phys.\ Lett.\ B {\bf 550}, 99 (2002)}
  %doi:10.1016/S0370-2693(02)02956-8
  [\href{https://arxiv.org/abs/hep-ph/0205336}{hep-ph/0205336}];
  %%CITATION = doi:10.1016/S0370-2693(02)02956-8;%%
  %57 citations counted in INSPIRE as of 02 Oct 2017
  N.~Maekawa and Y.~Muramatsu,
  %``Flavor changing nucleon decay,''
  \href{https://doi.org/10.1016/j.physletb.2017.02.028}{Phys.\ Lett.\ B {\bf 767}, 398 (2017)}
  %doi:10.1016/j.physletb.2017.02.028
  [\href{https://arxiv.org/abs/1601.04789}{arXiv:1601.04789 [hep-ph]}].
  %%CITATION = doi:10.1016/j.physletb.2017.02.028;%%
  %1 citations counted in INSPIRE as of 02 Oct 2017

\bibitem{minSO10}
%\bibitem{Georgi:1974my} 
  H.~Georgi,
  %``The State of the Art—Gauge Theories,''
  \href{https://doi.org/10.1063/1.2947450}{AIP Conf.\ Proc.\  {\bf 23}, 575 (1975)};
  %doi:10.1063/1.2947450
  %%CITATION = doi:10.1063/1.2947450;%%
  %257 citations counted in INSPIRE as of 23 Nov 2019
%\bibitem{Fritzsch:1974nn}
  H.~Fritzsch and P.~Minkowski,
  %``Unified Interactions of Leptons and Hadrons,''
  \href{https://doi.org/10.1016/0003-4916(75)90211-0}{Annals Phys.\  {\bf 93}, 193 (1975)}.
  %doi:10.1016/0003-4916(75)90211-0
  %%CITATION = doi:10.1016/0003-4916(75)90211-0;%%
  %1771 citations counted in INSPIRE as of 23 Nov 2019


\bibitem{seesawI}
  P.~Minkowski,
  %``$\mu \to e\gamma$ at a Rate of One Out of $10^{9}$ Muon Decays?,''
  \href{https://doi.org/10.1016/0370-2693(77)90435-X}{Phys.\ Lett.\  {\bf 67B}, 421 (1977)};
  %doi:10.1016/0370-2693(77)90435-X
  %%CITATION = doi:10.1016/0370-2693(77)90435-X;%%
  %2887 citations counted in INSPIRE as of 05 Oct 2017
  M.~Gell-Mann, P.~Ramond and R.~Slansky,
  %``Complex Spinors and Unified Theories,''
  Conf.\ Proc.\ C {\bf 790927}, 315 (1979)
  [\href{https://arxiv.org/abs/1306.4669}{arXiv:1306.4669 [hep-th]}];
  %%CITATION = ARXIV:1306.4669;%%
  %2559 citations counted in INSPIRE as of 05 Oct 2017
  T.~Yanagida,
  %``Horizontal Symmetry And Masses Of Neutrinos,''
  Conf.\ Proc.\ C {\bf 7902131}, 95 (1979);
  %%CITATION = CONFP,C7902131,95;%%
  %1361 citations counted in INSPIRE as of 05 Oct 2017
  S.~L.~Glashow,
  %``The Future of Elementary Particle Physics,''
  \href{https://doi.org/10.1007/978-1-4684-7197-7_15}{NATO Sci.\ Ser.\ B {\bf 61}, 687 (1980)};
  %doi:10.1007/978-1-4684-7197-7_15
  %%CITATION = doi:10.1007/978-1-4684-7197-7_15;%%
  %305 citations counted in INSPIRE as of 05 Oct 2017
  R.~N.~Mohapatra and G.~Senjanovic,
  %``Neutrino Mass and Spontaneous Parity Violation,''
  \href{https://doi.org/10.1103/PhysRevLett.44.912}{Phys.\ Rev.\ Lett.\  {\bf 44}, 912 (1980)}.
  %doi:10.1103/PhysRevLett.44.912
  %%CITATION = doi:10.1103/PhysRevLett.44.912;%%
  %4445 citations counted in INSPIRE as of 05 Oct 2017

\bibitem{seesawII}
  W.~Konetschny and W.~Kummer,
  %``Nonconservation of Total Lepton Number with Scalar Bosons,''
  \href{https://doi.org/10.1016/0370-2693(77)90407-5}{Phys.\ Lett.\  {\bf 70B}, 433 (1977)};
  %doi:10.1016/0370-2693(77)90407-5
  %%CITATION = doi:10.1016/0370-2693(77)90407-5;%%
  %271 citations counted in INSPIRE as of 05 Oct 2017
  T.~P.~Cheng and L.~F.~Li,
  %``Neutrino Masses, Mixings and Oscillations in SU(2) x U(1) Models of Electroweak Interactions,''
  \href{https://doi.org/10.1103/PhysRevD.22.2860}{Phys.\ Rev.\ D {\bf 22}, 2860 (1980)};
  %doi:10.1103/PhysRevD.22.2860
  %%CITATION = doi:10.1103/PhysRevD.22.2860;%%
  %809 citations counted in INSPIRE as of 05 Oct 2017
  G.~Lazarides, Q.~Shafi and C.~Wetterich,
  %``Proton Lifetime and Fermion Masses in an SO(10) Model,''
  \href{https://doi.org/10.1016/0550-3213(81)90354-0}{Nucl.\ Phys.\ B {\bf 181}, 287 (1981)};
  %doi:10.1016/0550-3213(81)90354-0
  %%CITATION = doi:10.1016/0550-3213(81)90354-0;%%
  %1121 citations counted in INSPIRE as of 05 Oct 2017
  J.~Schechter and J.~W.~F.~Valle,
  %``Neutrino Masses in SU(2) x U(1) Theories,''
  \href{https://doi.org/10.1103/PhysRevD.22.2227}{Phys.\ Rev.\ D {\bf 22}, 2227 (1980)};
  %doi:10.1103/PhysRevD.22.2227
  %%CITATION = doi:10.1103/PhysRevD.22.2227;%%
  %2161 citations counted in INSPIRE as of 05 Oct 2017
  R.~N.~Mohapatra and G.~Senjanovic,
  %``Neutrino Masses and Mixings in Gauge Models with Spontaneous Parity Violation,''
  \href{https://doi.org/10.1103/PhysRevD.23.165}{Phys.\ Rev.\ D {\bf 23}, 165 (1981)}.
  %doi:10.1103/PhysRevD.23.165
  %%CITATION = doi:10.1103/PhysRevD.23.165;%%
  %2149 citations counted in INSPIRE as of 05 Oct 2017

\bibitem{SUSYthreshold}
  A.~L.~Kagan and C.~H.~Albright,
  %``Quark and Lepton Masses in Superstring Type Models with Mirror Families,''
  \href{https://doi.org/10.1103/PhysRevD.38.917}{Phys.\ Rev.\ D {\bf 38}, 917 (1988)};
  %doi:10.1103/PhysRevD.38.917
  %%CITATION = doi:10.1103/PhysRevD.38.917;%%
  %14 citations counted in INSPIRE as of 12 Oct 2017
  T.~Banks,
  %``Supersymmetry and the Quark Mass Matrix,''
  \href{https://doi.org/10.1016/0550-3213(88)90222-2}{Nucl.\ Phys.\ B {\bf 303}, 172 (1988)};
  %doi:10.1016/0550-3213(88)90222-2
  %%CITATION = doi:10.1016/0550-3213(88)90222-2;%%
  %191 citations counted in INSPIRE as of 12 Oct 2017
  E.~Ma,
  %``Radiative Quark and Lepton Masses Through Soft Supersymmetry Breaking,''
  \href{https://doi.org/10.1103/PhysRevD.39.1922}{Phys.\ Rev.\ D {\bf 39}, 1922 (1989)}.
  %doi:10.1103/PhysRevD.39.1922
  %%CITATION = doi:10.1103/PhysRevD.39.1922;%%
  %58 citations counted in INSPIRE as of 12 Oct 2017
 
\bibitem{YU_SUSYthreshold}
  R.~Hempfling,
  %``Yukawa coupling unification with supersymmetric threshold corrections,''
  \href{https://doi.org/10.1103/PhysRevD.49.6168}{Phys.\ Rev.\ D {\bf 49}, 6168 (1994)};
  %doi:10.1103/PhysRevD.49.6168
  %%CITATION = doi:10.1103/PhysRevD.49.6168;%%
  %472 citations counted in INSPIRE as of 01 Aug 2019
  L.~J.~Hall, R.~Rattazzi and U.~Sarid,
  %``The Top quark mass in supersymmetric SO(10) unification,''
  \href{https://doi.org/10.1103/PhysRevD.50.7048}{Phys.\ Rev.\ D {\bf 50}, 7048 (1994)}
  %doi:10.1103/PhysRevD.50.7048
  [\href{https://arxiv.org/abs/hep-ph/9306309}{hep-ph/9306309}];
  %%CITATION = doi:10.1103/PhysRevD.50.7048;%%
  %950 citations counted in INSPIRE as of 01 Aug 2019
  M.~Carena, M.~Olechowski, S.~Pokorski and C.~E.~M.~Wagner,
  %``Electroweak symmetry breaking and bottom - top Yukawa unification,''
  \href{https://doi.org/10.1016/0550-3213(94)90313-1}{Nucl.\ Phys.\ B {\bf 426}, 269 (1994)}
  %doi:10.1016/0550-3213(94)90313-1
  [\href{https://arxiv.org/abs/hep-ph/9402253}{hep-ph/9402253}];
  %%CITATION = doi:10.1016/0550-3213(94)90313-1;%%
  %811 citations counted in INSPIRE as of 01 Aug 2019
  H.~Baer and J.~Ferrandis,
  %``Supersymmetric SO(10) GUT models with Yukawa unification and a positive mu term,''
  \href{https://doi.org/10.1103/PhysRevLett.87.211803}{Phys.\ Rev.\ Lett.\  {\bf 87}, 211803 (2001)}
  %doi:10.1103/PhysRevLett.87.211803
  [\href{https://arxiv.org/abs/hep-ph/0106352}{hep-ph/0106352}];
  %%CITATION = doi:10.1103/PhysRevLett.87.211803;%%
  %91 citations counted in INSPIRE as of 01 Aug 2019
  T.~Blazek, R.~Dermisek and S.~Raby,
  %``Predictions for Higgs and supersymmetry spectra from SO(10) Yukawa unification with mu greater than 0,''
  \href{https://doi.org/10.1103/PhysRevLett.88.111804}{Phys.\ Rev.\ Lett.\  {\bf 88}, 111804 (2002)}
  %doi:10.1103/PhysRevLett.88.111804
  [\href{https://arxiv.org/abs/hep-ph/0107097}{hep-ph/0107097}];
  %%CITATION = doi:10.1103/PhysRevLett.88.111804;%%
  %174 citations counted in INSPIRE as of 01 Aug 2019
  K.~Tobe and J.~D.~Wells,
  %``Revisiting top bottom tau Yukawa unification in supersymmetric grand unified theories,''
  \href{https://doi.org/10.1016/S0550-3213(03)00373-0}{Nucl.\ Phys.\ B {\bf 663}, 123 (2003)}
  %doi:10.1016/S0550-3213(03)00373-0
  [\href{https://arxiv.org/abs/hep-ph/0301015}{hep-ph/0301015}];
  %%CITATION = doi:10.1016/S0550-3213(03)00373-0;%%
  %100 citations counted in INSPIRE as of 01 Aug 2019
  D.~Auto, H.~Baer, C.~Balazs, A.~Belyaev, J.~Ferrandis and X.~Tata,
  %``Yukawa coupling unification in supersymmetric models,''
  \href{https://doi.org/10.1088/1126-6708/2003/06/023}{JHEP {\bf 0306}, 023 (2003)}
  %doi:10.1088/1126-6708/2003/06/023
  [\href{https://arxiv.org/abs/hep-ph/0302155}{hep-ph/0302155}];
  %%CITATION = doi:10.1088/1126-6708/2003/06/023;%%
  %135 citations counted in INSPIRE as of 01 Aug 2019
  M.~Albrecht, W.~Altmannshofer, A.~J.~Buras, D.~Guadagnoli and D.~M.~Straub,
  %``Challenging $SO(10)$ SUSY GUTs with family symmetries through FCNC processes,''
  \href{https://doi.org/10.1088/1126-6708/2007/10/055}{JHEP {\bf 0710}, 055 (2007)}
  %doi:10.1088/1126-6708/2007/10/055
  [\href{https://arxiv.org/abs/0707.3954}{arXiv:0707.3954 [hep-ph]}];
  %%CITATION = doi:10.1088/1126-6708/2007/10/055;%%
  %52 citations counted in INSPIRE as of 01 Aug 2019
%%%%%% add 191206 %%%%%%
  S.~Antusch and M.~Spinrath,
  %``Quark and lepton masses at the GUT scale including SUSY threshold corrections,''
  \href{https://doi.org/10.1103/PhysRevD.78.075020}{Phys.\ Rev.\ D {\bf 78}, 075020 (2008)}
  %doi:10.1103/PhysRevD.78.075020
  [\href{https://arxiv.org/abs/0804.0717}{arXiv:0804.0717 [hep-ph]}];
  %%CITATION = doi:10.1103/PhysRevD.78.075020;%%
  %93 citations counted in INSPIRE as of 03 Dec 2019
%%%%%% add 191206 %%%%%%
  S.~Antusch and M.~Spinrath,
  %``New GUT predictions for quark and lepton mass ratios confronted with phenomenology,''
  \href{https://doi.org/10.1103/PhysRevD.79.095004}{Phys.\ Rev.\ D {\bf 79}, 095004 (2009)}
  %doi:10.1103/PhysRevD.79.095004
  [\href{https://arxiv.org/abs/0902.4644}{arXiv:0902.4644 [hep-ph]}];
  %%CITATION = doi:10.1103/PhysRevD.79.095004;%%
  %100 citations counted in INSPIRE as of 03 Dec 2019
%%%%%% add 191206 %%%%%%
  A.~Crivellin and C.~Greub,
  %``Two-loop supersymmetric QCD corrections to Higgs-quark-quark couplings in the generic MSSM,''
  \href{https://doi.org/10.1103/PhysRevD.87.015013}{Phys.\ Rev.\ D {\bf 87}, 015013 (2013)}
  [Erratum: \href{https://doi.org/10.1103/PhysRevD.87.079901}{Phys.\ Rev.\ D {\bf 87}, 079901 (2013)}]
  %doi:10.1103/PhysRevD.87.015013, 10.1103/PhysRevD.87.079901
  [\href{https://arxiv.org/abs/1210.7453}{arXiv:1210.7453 [hep-ph]}].
  %%CITATION = doi:10.1103/PhysRevD.87.015013, 10.1103/PhysRevD.87.079901;%%
  %40 citations counted in INSPIRE as of 03 Dec 2019
  A.~Anandakrishnan, S.~Raby and A.~Wingerter,
  %``Yukawa Unification Predictions for the LHC,''
  \href{https://doi.org/10.1103/PhysRevD.87.055005}{Phys.\ Rev.\ D {\bf 87}, 055005 (2013)}
  %doi:10.1103/PhysRevD.87.055005
  [\href{https://arxiv.org/abs/1212.0542}{arXiv:1212.0542 [hep-ph]}].
  %%CITATION = doi:10.1103/PhysRevD.87.055005;%%
  %38 citations counted in INSPIRE as of 01 Aug 2019


\bibitem{SUSYthresholdCKM}
  T.~Blazek, S.~Raby and S.~Pokorski,
  %``Finite supersymmetric threshold corrections to CKM matrix elements in the large tan Beta regime,''
  \href{https://doi.org/10.1103/PhysRevD.52.4151}{Phys.\ Rev.\ D {\bf 52}, 4151 (1995)}
  %doi:10.1103/PhysRevD.52.4151
  [\href{https://arxiv.org/abs/hep-ph/9504364}{hep-ph/9504364}];
  %%CITATION = doi:10.1103/PhysRevD.52.4151;%%
  %217 citations counted in INSPIRE as of 12 Oct 2017
  J.~Ferrandis and N.~Haba,
  %``Supersymmetry breaking as the origin of flavor,''
  \href{https://doi.org/10.1103/PhysRevD.70.055003}{Phys.\ Rev.\ D {\bf 70}, 055003 (2004)}
  %doi:10.1103/PhysRevD.70.055003
  [\href{https://arxiv.org/abs/hep-ph/0404077}{hep-ph/0404077}];
  %%CITATION = doi:10.1103/PhysRevD.70.055003;%%
  %28 citations counted in INSPIRE as of 12 Oct 2017
  A.~Crivellin and U.~Nierste,
  %``Supersymmetric renormalisation of the CKM matrix and new constraints on the squark mass matrices,''
  \href{https://doi.org/10.1103/PhysRevD.79.035018}{Phys.\ Rev.\ D {\bf 79}, 035018 (2009)}
  %doi:10.1103/PhysRevD.79.035018
  [\href{https://arxiv.org/abs/0810.1613}{arXiv:0810.1613 [hep-ph]}];
  %%CITATION = doi:10.1103/PhysRevD.79.035018;%%
  %72 citations counted in INSPIRE as of 12 Oct 2017
%%%%%% add 191206 %%%%%%
  A.~Crivellin, L.~Hofer, U.~Nierste and D.~Scherer,
  %``Phenomenological consequences of radiative flavor violation in the MSSM,''
  \href{https://doi.org/10.1103/PhysRevD.84.035030}{Phys.\ Rev.\ D {\bf 84}, 035030 (2011)}
  %doi:10.1103/PhysRevD.84.035030
  [\href{https://arxiv.org/abs/1105.2818}{arXiv:1105.2818 [hep-ph]}].
  %%CITATION = doi:10.1103/PhysRevD.84.035030;%%
  %45 citations counted in INSPIRE as of 03 Dec 2019

%%%%%% add 191206 %%%%%%
\bibitem{Crivellin:2009ar} 
  A.~Crivellin and U.~Nierste,
  %``Chirally enhanced corrections to FCNC processes in the generic MSSM,''
  \href{https://doi.org/10.1103/PhysRevD.81.095007}{Phys.\ Rev.\ D {\bf 81}, 095007 (2010)}
  %doi:10.1103/PhysRevD.81.095007
  [\href{https://arxiv.org/abs/0908.4404}{arXiv:0908.4404 [hep-ph]}].
  %%CITATION = doi:10.1103/PhysRevD.81.095007;%%
  %43 citations counted in INSPIRE as of 03 Dec 2019

\bibitem{Crivellin:2011jt} 
  A.~Crivellin, L.~Hofer and J.~Rosiek,
  %``Complete resummation of chirally-enhanced loop-effects in the MSSM with non-minimal sources of flavor-violation,''
  \href{https://doi.org/10.1007/JHEP07(2011)017}{JHEP {\bf 1107}, 017 (2011)}
  %doi:10.1007/JHEP07(2011)017
  [\href{https://arxiv.org/abs/1103.4272}{arXiv:1103.4272 [hep-ph]}].
  %%CITATION = doi:10.1007/JHEP07(2011)017;%%
  %60 citations counted in INSPIRE as of 16 Oct 2017
  
\bibitem{CCB}
  M.~Drees, M.~Gluck and K.~Grassie,
  %``A New Class of False Vacua in Low-energy $N=1$ Supergravity Theories,''
  \href{https://doi.org/10.1016/0370-2693(85)91538-2}{Phys.\ Lett.\  {\bf 157B}, 164 (1985)};
  %doi:10.1016/0370-2693(85)91538-2
  %%CITATION = doi:10.1016/0370-2693(85)91538-2;%%
  %88 citations counted in INSPIRE as of 14 Apr 2019
  J.~F.~Gunion, H.~E.~Haber and M.~Sher,
  %``Charge / Color Breaking Minima and a-Parameter Bounds in Supersymmetric Models,''
  \href{https://doi.org/10.1016/0550-3213(88)90168-X}{Nucl.\ Phys.\ B {\bf 306}, 1 (1988)};
  %doi:10.1016/0550-3213(88)90168-X
  %%CITATION = doi:10.1016/0550-3213(88)90168-X;%%
  %195 citations counted in INSPIRE as of 14 Apr 2019
  H.~Komatsu,
  %``New Constraints on Parameters in the Minimal Supersymmetric Model,''
  \href{https://doi.org/10.1016/0370-2693(88)91441-4}{Phys.\ Lett.\ B {\bf 215}, 323 (1988)};
  %doi:10.1016/0370-2693(88)91441-4
  %%CITATION = doi:10.1016/0370-2693(88)91441-4;%%
  %95 citations counted in INSPIRE as of 14 Apr 2019
  P.~Langacker and N.~Polonsky,
  %``Implications of Yukawa unification for the Higgs sector in supersymmetric grand unified models,''
  \href{https://doi.org/10.1103/PhysRevD.50.2199}{Phys.\ Rev.\ D {\bf 50}, 2199 (1994)}
  %doi:10.1103/PhysRevD.50.2199
  [\href{https://arxiv.org/abs/hep-ph/9403306}{hep-ph/9403306}];
  %%CITATION = doi:10.1103/PhysRevD.50.2199;%%
  %135 citations counted in INSPIRE as of 14 Apr 2019
  J.~A.~Casas, A.~Lleyda and C.~Munoz,
  %``Strong constraints on the parameter space of the MSSM from charge and color breaking minima,''
  \href{https://doi.org/10.1016/0550-3213(96)00194-0}{Nucl.\ Phys.\ B {\bf 471}, 3 (1996)}
  %doi:10.1016/0550-3213(96)00194-0
  [\href{https://arxiv.org/abs/hep-ph/9507294}{hep-ph/9507294}];
  %%CITATION = doi:10.1016/0550-3213(96)00194-0;%%
  %361 citations counted in INSPIRE as of 14 Apr 2019
  J.~A.~Casas and S.~Dimopoulos,
  %``Stability bounds on flavor violating trilinear soft terms in the MSSM,''
  \href{https://doi.org/10.1016/0370-2693(96)01000-3}{Phys.\ Lett.\ B {\bf 387}, 107 (1996)}
  %doi:10.1016/0370-2693(96)01000-3
  [\href{https://arxiv.org/abs/hep-ph/9606237}{hep-ph/9606237}].
  %%CITATION = doi:10.1016/0370-2693(96)01000-3;%%
  %196 citations counted in INSPIRE as of 14 Apr 2019 

\bibitem{Ellis:1979fg} 
  J.~R.~Ellis and M.~K.~Gaillard,
  %``Fermion Masses and Higgs Representations in SU(5),''
  \href{https://doi.org/10.1016/0370-2693(79)90476-3}{Phys.\ Lett.\  {\bf 88B}, 315 (1979)}.
  %doi:10.1016/0370-2693(79)90476-3
  %%CITATION = doi:10.1016/0370-2693(79)90476-3;%%
  %230 citations counted in INSPIRE as of 11 Apr 2019

\bibitem{PDG}
  M.~Tanabashi {\it et al.} [Particle Data Group],
  %``Review of Particle Physics,''
  \href{https://doi.org/10.1103/PhysRevD.98.030001}{Phys.\ Rev.\ D {\bf 98}, 030001 (2018)}.
  %doi:10.1103/PhysRevD.98.030001
  %%CITATION = doi:10.1103/PhysRevD.98.030001;%%
  %1495 citations counted in INSPIRE as of 27 Apr 2019

\bibitem{RunDec}
  K.~G.~Chetyrkin, J.~H.~Kuhn and M.~Steinhauser,
  %``RunDec: A Mathematica package for running and decoupling of the strong coupling and quark masses,''
  \href{https://doi.org/10.1016/S0010-4655(00)00155-7}{Comput.\ Phys.\ Commun.\  {\bf 133}, 43 (2000)}
  %doi:10.1016/S0010-4655(00)00155-7
  [\href{https://arxiv.org/abs/hep-ph/0004189}{hep-ph/0004189}].
  %%CITATION = doi:10.1016/S0010-4655(00)00155-7;%%
  %245 citations counted in INSPIRE as of 01 f\UTF{00E9}vr. 2016

\bibitem{SMrun}
  H.~Arason, D.~J.~Castano, B.~Keszthelyi, S.~Mikaelian, E.~J.~Piard, P.~Ramond and B.~D.~Wright,
  %``Renormalization group study of the standard model and its extensions. 1. The Standard model,''
  \href{https://doi.org/10.1103/PhysRevD.46.3945}{Phys.\ Rev.\ D {\bf 46}, 3945 (1992)}.
  %doi:10.1103/PhysRevD.46.3945
  %%CITATION = doi:10.1103/PhysRevD.46.3945;%%
  %357 citations counted in INSPIRE as of 01 f\UTF{00E9}vr. 2016

\bibitem{Machacek_Vaughn} 
  M.~E.~Machacek and M.~T.~Vaughn,
  %``Two Loop Renormalization Group Equations in a General Quantum Field Theory. 1. Wave Function Renormalization,''
  \href{https://doi.org/10.1016/0550-3213(83)90610-7}{Nucl.\ Phys.\ B {\bf 222}, 83 (1983)};
  %doi:10.1016/0550-3213(83)90610-7
  %%CITATION = doi:10.1016/0550-3213(83)90610-7;%%
  %576 citations counted in INSPIRE as of 05 Dec 2017
  M.~E.~Machacek and M.~T.~Vaughn,
  %``Two Loop Renormalization Group Equations in a General Quantum Field Theory. 2. Yukawa Couplings,''
  \href{https://doi.org/10.1016/0550-3213(84)90533-9}{Nucl.\ Phys.\ B {\bf 236}, 221 (1984)};
  %doi:10.1016/0550-3213(84)90533-9
  %%CITATION = doi:10.1016/0550-3213(84)90533-9;%%
  %509 citations counted in INSPIRE as of 05 Dec 2017
  M.~E.~Machacek and M.~T.~Vaughn,
  %``Two Loop Renormalization Group Equations in a General Quantum Field Theory. 3. Scalar Quartic Couplings,''
  \href{https://doi.org/10.1016/0550-3213(85)90040-9}{Nucl.\ Phys.\ B {\bf 249}, 70 (1985)}.
  %doi:10.1016/0550-3213(85)90040-9
  %%CITATION = doi:10.1016/0550-3213(85)90040-9;%%
  %417 citations counted in INSPIRE as of 05 Dec 2017

\bibitem{Luo:2002ey} 
  M.~x.~Luo and Y.~Xiao,
  %``Two loop renormalization group equations in the standard model,''
  \href{https://doi.org/10.1103/PhysRevLett.90.011601}{Phys.\ Rev.\ Lett.\  {\bf 90}, 011601 (2003)}
  %doi:10.1103/PhysRevLett.90.011601
  [\href{https://arxiv.org/abs/hep-ph/0207271}{hep-ph/0207271}].
  %%CITATION = doi:10.1103/PhysRevLett.90.011601;%%
  %112 citations counted in INSPIRE as of 05 Dec 2017

\bibitem{Bednyakov:2012en} 
  A.~V.~Bednyakov, A.~F.~Pikelner and V.~N.~Velizhanin,
  %``Yukawa coupling beta-functions in the Standard Model at three loops,''
  \href{https://doi.org/10.1016/j.physletb.2013.04.038}{Phys.\ Lett.\ B {\bf 722}, 336 (2013)}
  %doi:10.1016/j.physletb.2013.04.038
  [\href{https://arxiv.org/abs/1212.6829}{arXiv:1212.6829 [hep-ph]}].
  %%CITATION = doi:10.1016/j.physletb.2013.04.038;%%
  %70 citations counted in INSPIRE as of 05 Dec 2017

\bibitem{Chetyrkin:2013wya} 
  K.~G.~Chetyrkin and M.~F.~Zoller,
  %``$\beta$-function for the Higgs self-interaction in the Standard Model at three-loop level,''
  \href{https://doi.org/10.1007/JHEP04(2013)091}{JHEP {\bf 1304}, 091 (2013)}
  [Erratum: \href{https://doi.org/10.1007/JHEP09(2013)155}{JHEP {\bf 1309}, 155 (2013)}]
  %doi:10.1007/JHEP04(2013)091, 10.1007/JHEP09(2013)155
  [\href{https://arxiv.org/abs/1303.2890}{arXiv:1303.2890 [hep-ph]}].
  %%CITATION = doi:10.1007/JHEP04(2013)091, 10.1007/JHEP09(2013)155;%%
  %90 citations counted in INSPIRE as of 05 Dec 2017

%%%%%% add 191206 %%%%%%
\bibitem{Antusch:2013jca} 
  S.~Antusch and V.~Maurer,
  %``Running quark and lepton parameters at various scales,''
  \href{https://doi.org/10.1007/JHEP11(2013)115}{JHEP {\bf 1311}, 115 (2013)}
  %doi:10.1007/JHEP11(2013)115
  [\href{https://arxiv.org/abs/1306.6879}{arXiv:1306.6879 [hep-ph]}];
  %%CITATION = doi:10.1007/JHEP11(2013)115;%%
  %81 citations counted in INSPIRE as of 03 Dec 2019
  V.~K.~M.~Maurer,
  %``Insights into Grand Unified Theories from Current Experimental Data,''
  \href{https://doi.org/10.5451/unibas-006387307}{doi:10.5451/unibas-006387307}
  %%CITATION = doi:10.5451/unibas-006387307;%%

\bibitem{MStoDR}
  S.~P.~Martin and M.~T.~Vaughn,
  %``Regularization dependence of running couplings in softly broken supersymmetry,''
  \href{https://doi.org/10.1016/0370-2693(93)90136-6}{Phys.\ Lett.\ B {\bf 318}, 331 (1993)}
  %doi:10.1016/0370-2693(93)90136-6
  [\href{https://arxiv.org/abs/hep-ph/9308222}{hep-ph/9308222}].
  %%CITATION = doi:10.1016/0370-2693(93)90136-6;%%
  %247 citations counted in INSPIRE as of 27 Feb 2016

\bibitem{MSSMrun}
  S.~P.~Martin and M.~T.~Vaughn,
  %``Two loop renormalization group equations for soft supersymmetry breaking couplings,''
  \href{https://doi.org/10.1103/PhysRevD.50.2282}{Phys.\ Rev.\ D {\bf 50}, 2282 (1994)}
  [Erratum: \href{https://doi.org/10.1103/PhysRevD.78.039903}{Phys.\ Rev.\ D {\bf 78}, 039903 (2008)}]
  %doi:10.1103/PhysRevD.50.2282, 10.1103/PhysRevD.78.039903
  [\href{https://arxiv.org/abs/hep-ph/9311340}{hep-ph/9311340}].
  %%CITATION = doi:10.1103/PhysRevD.50.2282, 10.1103/PhysRevD.78.039903;%%
  %788 citations counted in INSPIRE as of 27 Apr 2019
  
\bibitem{large_tanbeta} 
  G.~F.~Giudice and G.~Ridolfi,
  %``Constraints on the Minimal $N=1$ Supergravity Theory From Electroweak Symmetry Breaking,''
  \href{https://doi.org/10.1007/BF01585630}{Z.\ Phys.\ C {\bf 41}, 447 (1988)};
  %doi:10.1007/BF01585630
  %%CITATION = doi:10.1007/BF01585630;%%
  %103 citations counted in INSPIRE as of 24 Oct 2017
  M.~Olechowski and S.~Pokorski,
  %``Hierarchy of Quark Masses in the Isotopic Doublets in N=1 Supergravity Models,''
  \href{https://doi.org/10.1016/0370-2693(88)91383-4}{Phys.\ Lett.\ B {\bf 214}, 393 (1988)};
  %doi:10.1016/0370-2693(88)91383-4
  %%CITATION = doi:10.1016/0370-2693(88)91383-4;%%
  %171 citations counted in INSPIRE as of 24 Oct 2017
  B.~Ananthanarayan, G.~Lazarides and Q.~Shafi,
  %``Top mass prediction from supersymmetric guts,''
  \href{https://doi.org/10.1103/PhysRevD.44.1613}{Phys.\ Rev.\ D {\bf 44}, 1613 (1991)};
  %doi:10.1103/PhysRevD.44.1613
  %%CITATION = doi:10.1103/PhysRevD.44.1613;%%
  %317 citations counted in INSPIRE as of 24 Oct 2017
  H.~Arason, D.~Castano, B.~Keszthelyi, S.~Mikaelian, E.~Piard, P.~Ramond and B.~Wright,
  %``Top quark and Higgs mass bounds from a numerical study of superGUTs,''
  \href{https://doi.org/10.1103/PhysRevLett.67.2933}{Phys.\ Rev.\ Lett.\  {\bf 67}, 2933 (1991)};
  %doi:10.1103/PhysRevLett.67.2933
  %%CITATION = doi:10.1103/PhysRevLett.67.2933;%%
  %227 citations counted in INSPIRE as of 24 Oct 2017
  S.~Kelley, J.~L.~Lopez and D.~V.~Nanopoulos,
  %``Yukawa unification,''
  \href{https://doi.org/10.1016/0370-2693(92)92003-Y}{Phys.\ Lett.\ B {\bf 274}, 387 (1992)}.
  %doi:10.1016/0370-2693(92)92003-Y
  %%CITATION = doi:10.1016/0370-2693(92)92003-Y;%%
  %93 citations counted in INSPIRE as of 24 Oct 2017

\bibitem{Gabbiani:1996hi} 
  F.~Gabbiani, E.~Gabrielli, A.~Masiero and L.~Silvestrini,
  %``A Complete analysis of FCNC and CP constraints in general SUSY extensions of the standard model,''
  \href{https://doi.org/10.1016/0550-3213(96)00390-2}{Nucl.\ Phys.\ B {\bf 477}, 321 (1996)}
  %doi:10.1016/0550-3213(96)00390-2
  [\href{https://arxiv.org/abs/hep-ph/9604387}{hep-ph/9604387}].
  %%CITATION = doi:10.1016/0550-3213(96)00390-2;%%
  %1335 citations counted in INSPIRE as of 21 Oct 2019

\bibitem{UTfit}
%\bibitem{Bona:2006sa} 
  M.~Bona {\it et al.} [UTfit Collaboration],
  %``Constraints on new physics from the quark mixing unitarity triangle,''
  \href{https://doi.org/10.1103/PhysRevLett.97.151803}{Phys.\ Rev.\ Lett.\  {\bf 97}, 151803 (2006)}
  %doi:10.1103/PhysRevLett.97.151803
  [\href{https://arxiv.org/abs/hep-ph/0605213}{hep-ph/0605213}];
  %%CITATION = doi:10.1103/PhysRevLett.97.151803;%%
  %167 citations counted in INSPIRE as of 21 Oct 2019
%\bibitem{Bona:2007vi} 
  M.~Bona {\it et al.} [UTfit Collaboration],
  %``Model-independent constraints on $\Delta F=2$ operators and the scale of new physics,''
  \href{https://doi.org/10.1088/1126-6708/2008/03/049}{JHEP {\bf 0803}, 049 (2008)}
  %doi:10.1088/1126-6708/2008/03/049
  [\href{https://arxiv.org/abs/0707.0636}{arXiv:0707.0636 [hep-ph]}];
  %%CITATION = doi:10.1088/1126-6708/2008/03/049;%%
  %486 citations counted in INSPIRE as of 21 Oct 2019
  %see also http://www.utfit.org/UTfit/WebHome for the latest results.

\bibitem{Paradisi:2005fk} 
  P.~Paradisi,
  %``Constraints on SUSY lepton flavor violation by rare processes,''
  \href{https://doi.org/10.1088/1126-6708/2005/10/006}{JHEP {\bf 0510}, 006 (2005)}
  %doi:10.1088/1126-6708/2005/10/006
  [\href{https://arxiv.org/abs/hep-ph/0505046}{hep-ph/0505046}];
  %%CITATION = doi:10.1088/1126-6708/2005/10/006;%%
  %132 citations counted in INSPIRE as of 11 Jul 2018
%\bibitem{Arana-Catania:2013ggc} 
  M.~Arana-Catania, S.~Heinemeyer and M.~J.~Herrero,
  %``New Constraints on General Slepton Flavor Mixing,''
  \href{https://doi.org/10.1103/PhysRevD.88.015026}{Phys.\ Rev.\ D {\bf 88}, 015026 (2013)}
  %doi:10.1103/PhysRevD.88.015026
  [\href{https://arxiv.org/abs/1304.2783}{arXiv:1304.2783 [hep-ph]}].
  %%CITATION = doi:10.1103/PhysRevD.88.015026;%%
  %39 citations counted in INSPIRE as of 11 Oct 2019

\bibitem{Inoue:1982ej} 
  K.~Inoue, A.~Kakuto, H.~Komatsu and S.~Takeshita,
  %``Low-Energy Parameters and Particle Masses in a Supersymmetric Grand Unified Model,''
  \href{https://doi.org/10.1143/PTP.67.1889}{Prog.\ Theor.\ Phys.\  {\bf 67}, 1889 (1982)}.
  %doi:10.1143/PTP.67.1889
  %%CITATION = doi:10.1143/PTP.67.1889;%%
  %454 citations counted in INSPIRE as of 20 Apr 2019







%%%%%%%%%%%%%%
% Do not cite in the paper %
%%%%%%%%%%%%%%


%\bibitem{Ciuchini:2007cw}
%  M.~Ciuchini, E.~Franco, D.~Guadagnoli, V.~Lubicz, M.~Pierini, V.~Porretti and L.~Silvestrini,
  %``$D$ - $\bar{D}$ mixing and new physics: General considerations and constraints on the MSSM,''
%  Phys.\ Lett.\ B {\bf 655} (2007) 162
  %doi:10.1016/j.physletb.2007.08.055
%  [hep-ph/0703204].
  %%CITATION = doi:10.1016/j.physletb.2007.08.055;%%
  %140 citations counted in INSPIRE as of 04 Sep 2017
  
%\bibitem{Carrasco:2014uya} 
%  N.~Carrasco {\it et al.},
  %``$D^0$$-\bar{D}^0$ mixing in the standard model and beyond from $N_f$ =2 twisted mass QCD,''
%  Phys.\ Rev.\ D {\bf 90}, no. 1, 014502 (2014)
  %doi:10.1103/PhysRevD.90.014502
%  [arXiv:1403.7302 [hep-lat]].
  %%CITATION = doi:10.1103/PhysRevD.90.014502;%%
  %26 citations counted in INSPIRE as of 04 Sep 2017

%\bibitem{Masiero:2005ua} 
%  A.~Masiero, S.~K.~Vempati and O.~Vives,
  %``Flavour physics and grand unification,''
%  arXiv:0711.2903 [hep-ph].
  %%CITATION = ARXIV:0711.2903;%%
  %42 citations counted in INSPIRE as of 05 Sep 2017


\end{thebibliography}
\end{document}